# Lonely individuals process the world in idiosyncratic ways

Elisa C. Baek*[1], Ryan Hyon[1], Karina López[1], Meng Du[1], Mason A. Porter[2,3], and Carolyn Parkinson*[1,4]

[1]Department of Psychology, University of California, Los Angeles, [2] Department of Mathematics, University of California, Los Angeles, [3]Sante Fe Institute, [4] Brain Research Institute, University of California, Los Angeles

* Corresponding authors

**Abstract**

Loneliness is detrimental to well-being and is often accompanied by self-reported feelings of not being understood by others. What contributes to such feelings in lonely people? We used functional magnetic resonance imaging (fMRI) of 66 participants to unobtrusively measure the relative alignment of people's mental processing of naturalistic stimuli and tested whether or not lonely people actually process the world in idiosyncratic ways. We found evidence for such idiosyncrasy: lonely individuals' neural responses were dissimilar to their peers, particularly in regions of the default-mode network in which similar responses have been associated with shared perspectives and subjective understanding. These relationships persisted when controlling for demographic similarities, objective social isolation, and participants' friendships with each other. Our findings suggest the possibility that being surrounded by people who see the world differently from oneself, even if one is friends with them, may be a risk factor for loneliness.

**Statement of Relevance**

Loneliness (i.e., the distressing feeling that often accompanies subjective perceptions of social disconnection), which is pervasive and consequential, can have devastating consequences on both mental and physical well-being. Prior work suggests that lonely individuals self-report not feeling understood by others. Given the importance of feeling understood in achieving social connection, the feeling of not being understood by others may be one feature that characterizes loneliness. What contributes to such feelings in lonely individuals? In this paper, we used neuroimaging to show that lonely individuals process the world in idiosyncratic ways that are exceptionally dissimilar to their peers. Such idiosyncratic neural processing may contribute to feelings of disconnection from a lack of shared understanding. Our findings elucidate the underlying processes that characterize loneliness.

## Introduction

Humans have a fundamental need to belong and connect socially (Baumeister & Leary, 1995). When this need to belong is not met, there can be devastating consequences. It is well-established that loneliness (i.e., the distressing feeling that often accompanies subjective perceptions of social disconnection) has detrimental effects on the well-being of individuals, including increased risk for mortality that persists even after controlling for comorbidities (Cacioppo & Cacioppo, 2014; Hawkley et al., 2003; Hawkley & Cacioppo, 2010; Holt-Lunstad et al., 2010; Shankar et al., 2011).

Feeling understood by others is one critical factor for achieving social connection (Reis et al., 2000, 2017) and is associated with greater life satisfaction (Lun et al., 2008; Reis et al., 2000), more positive evaluations of interactions with strangers (Cross et al., 2000), and increased fulfillment in close relationships (Oishi et al., 2010). In one study, feeling understood activated brain regions that are associated with reward processing, whereas not feeling understood activated brain regions that are associated with negative affect (Morelli et al., 2013). Self-report data also suggest that there is an association between loneliness and not feeling understood by others (Cox et al., 2020; Routasalo et al., 2006). Such findings suggest that not feeling understood by others may be a risk factor for loneliness. However, it is unknown whether or not lonely people actually see the world in ways that are dissimilar to others in their community (rather than, e.g., exaggerating how dissimilar others' views are to their own), which may contribute to feelings of disconnection due to a lack of shared understanding.

We used neuroimaging to test the hypothesis that lonely[i] people have neural responses to naturalistic stimuli (specifically, videos) that are idiosyncratic in comparison to those of their peers (including other lonely people), perhaps contributing to the lack of feeling understood that often accompanies loneliness. Measuring brain responses during a naturalistic paradigm in which people view audiovisual stimuli that unfold over time provides a window into individuals' unconstrained thought processes as they develop and evolve (Sonkusare et al., 2019). Additionally, examining the similarity of neural responses can simultaneously capture various types of processing similarity, including similarities in high-level interpretations and understanding (Nguyen et al., 2019; Yeshurun et al., 2017, 2021), affective processing (Nummenmaa et al., 2012), and patterns of attention allocation (e.g., mind-wandering versus paying attention; Song et al., 2021; the aspects of stimuli to which people attend; Lahnakoski et al., 2014). Accordingly, the extent to which an individual exhibits similar neural responses to their peers can provide insights into the extent to which they process the world in a way that is similar to their peers.

[i] See Relating ISC with loneliness in the Method section for more details about how we classified participants as lonely.

Much work that uses inter-subject correlations (ISCs) of neural responses has looked at if and where similarities in behavior, interpretive frames and expectations, or traits are associated with similarities in neural responding (Leong et al., 2020; Nguyen et al., 2019; Regev et al., 2019; Yeshurun et al., 2017). Other approaches have calculated ISCs to examine how people's overall levels of particular traits or symptoms relate to their level of attunement with others (Bolis et al., 2017, 2021). Accordingly, we test the hypothesis that "Non-lonely people are all alike, but every lonely individual processes the world in their own idiosyncratic way." In other words, we test whether the associations between loneliness and neural responses to naturalistic stimuli follow an "Anna Karenina principle". This principle is inspired by the opening line from the novel *Anna Karenina*: "Happy families are all alike; every unhappy family is unhappy in its own way" (Tolstoy, 1997). It proposes that successful endeavors are marked by similar characteristics but that unsuccessful endeavors are each different in their own idiosyncratic way (Diamond, 1997). Studying various phenomena in light of this principle has yielded a variety of meaningful insights (Diamond, 1997; Finn et al., 2018; Zaneveld et al., 2017). In the present study, we tested whether individuals who are not lonely are exceptionally similar to each other in how they process the world, whereas lonely individuals each process the world in their own distinct way.

We focus on *subjective* social isolation (i.e., individuals' perceptions or feelings of isolation), but *objective* social isolation has also been linked to myriad negative health outcomes (Shankar et al., 2011). Objective social isolation is often measured by obtaining information about an individual's self-reported social-network size (Moieni & Eisenberger, 2020). Although objective and subjective social isolation are associated with each other, they are distinct constructs (Moieni & Eisenberger, 2020; Routasalo et al., 2006), and some evidence suggests that subjective feelings of social isolation are linked more strongly than objective social isolation to negative health outcomes (Holwerda et al., 2014; Lee & Ko, 2018). To disentangle people's perceptions of their objective levels of social connection/disconnection from their subjective feelings of social connection/disconnection, we conducted additional analyses that account for objective social isolation by controlling for the number of friends (i.e., out-degree centrality) that individuals reported having in their communities. This approach allowed us to test whether or not individuals who experience high levels of loneliness have neural responses that are dissimilar to those of their peers and to one another, even after controlling for their objective number of social ties. In other words, we test whether or not lonely individuals process the world idiosyncratically, even if they have many friends.

## Method

**fMRI study participants**. A total of 70 participants from two residential communities participated in our neuroimaging study, which consisted of a functional magnetic resonance (fMRI) scanning session and self-report questionnaires (see Fig. 1). We selected our sample size based on other neuroimaging studies with similar paradigms that relate neural similarity to behavioral traits (Finn et al., 2018; Nguyen et al., 2019; Parkinson et al., 2018). The participants in our study consisted of first-year students at a large public university in the United States. We excluded four participants from the fMRI data; two participants had excessive movement in more than half of the scan, one participant fell asleep during half of the scan, and one participant did not complete the scan. This resulted in a total of 66 participants (of whom 41 were female) between the ages of 18 and 21 (with a mean of 18.23 and a standard deviation of 0.63) that we included in our primary analyses of the relationship between ISCs and loneliness. Of these subjects, one participant had excessive head movement in one of the four runs and one participant reported falling asleep in one of the four runs. We excluded the respective runs for these participants and included only the remaining three runs for these participants in analyses that involved the brain data. All participants provided informed consent in accordance with the procedures of the Institutional Review Board of the university at which the study was conducted. We reported on separate analyses of the same data set in other papers to investigate associations between neural similarity and (1) aspects of individuals' positions in their social networks (Baek et al., 2022) and (2) personality similarity (Matz et al., 2021). In the present study, we investigate associations between neural similarity and individuals' subjective feelings of social disconnection.

**fMRI procedure.** Participants attended an in-person study session that included self-report surveys and a 90-minute neuroimaging session in which their brain activity was measured using blood-oxygen-level-dependent (BOLD) fMRI. The fMRI data collection occurred between September 2019 and early November 2019 during the participants' first year at the university and prior to the social-network survey (see Characterizing subjective and objective social disconnection). Prior to entering the scanner, participants completed self-report surveys in which they provided demographic information such as their age, gender, and race/ethnicity. During the fMRI portion of the study, the participants watched 14 different video clips with sound. These 14 videos ranged in duration (from 91 to 734 seconds) and content. (See Table S1 for descriptions of the content.) Prior to scanning, participants were informed that they would be watching video clips of heterogeneous content and that their experience would be akin to watching television while someone else "channel surfed"[ii]. We selected a subset of the video clips from ones that had been used previously, and we used similar criteria to those in prior works to select new stimuli (Hyon et al., 2020; Parkinson et al., 2018). First, in an effort to avoid inducing inter-subject differences from heterogeneous familiarities with content, we selected stimuli that were unlikely to have been seen before by our participants. Second, to minimize the likelihood that participants would engage in mind-wandering during viewing (as that could introduce undesirable noise into our data), we selected stimuli that were likely to be engaging. Third, we selected stimuli that were likely to elicit meaningful variability in interpretations and meaning that different individuals would draw from the content. Participants were asked to watch the videos naturally (i.e., as they would watch them in a normal situation in life). All participants saw the videos in the same order to avoid any potential variability in neural responses from differences in the way

[ii] The term "channel-surfing" is an idiom that refers to scanning through different television channels.

that the stimuli were presented (rather than from endogenous participant-level differences). The video "task" was divided into four runs, and each run consisted of a continuous stream of content. In each run, the video clips were presented immediately after one another, with no gap between the clips. The task lasted approximately 60 minutes in total. Structural images of the brain were also collected; see fMRI data acquisition for more detail.

**Characterizing subjective and objective social disconnection.** To characterize the participants' subjective feelings of social disconnection, we administered the UCLA Loneliness Scale (ULS-8) (Hays & Dimatteo, 1987) on the day of each participant's fMRI scan. We based our characterizations of participants' objective levels of social disconnection on their responses to a separate social-network survey that was administered during December and January of the students' first year at the university. (The academic year began at the end of September.) In the survey, participants were asked to type the names of other residents in their residential community with whom they interacted regularly. They were prompted with the following question: "Consider the people you like to spend your free time with. Since you arrived at [institution name], who are the people you've socialized with most often? (Examples: eat meals with, hang out with, study with, spend time with)." The participants were free to name as many people as they desired without any restrictions, and no time limit was imposed. This question was adapted from prior research that investigated the social networks of university students (Burt, 2004; Parkinson et al., 2017, 2018). Using the participants' answers to this question, we calculated out-degree centrality, which is equal to the number of the participant's community members with whom they reported socializing regularly. This allowed us to obtain an inverse measure of participants' objective levels of social disconnection (i.e., high out-degree centrality values reflect low levels of social disconnection and low out-degree centrality values reflect high levels of social disconnection), which we then used as a control variable.

A total of 119 subjects completed the social-network survey; 66 of these participants also participated in the fMRI session. Of these 66 participants who participated in both the social-network survey and the fMRI session, two participants were excluded due to excessive head movement and one participant was excluded due to falling asleep. This resulted in a total of 63 participants (40 of whom were female) that we included for all analyses that incorporated out-degree centrality as a control variable.

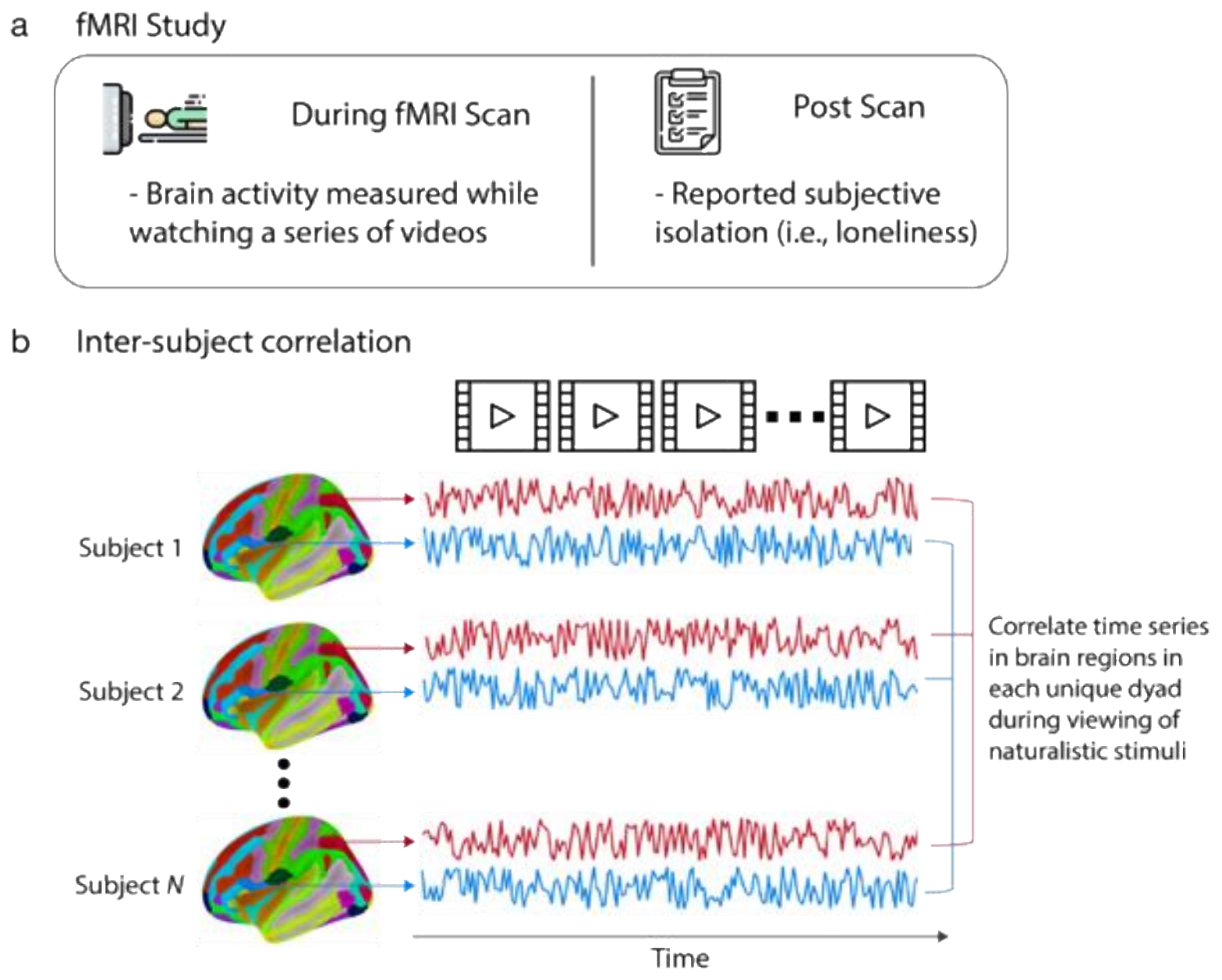

**Fig. 1.** Overview of the study paradigm and analysis. **(a)** Schematic of the study paradigm. Participants attended an in-lab session in which their brain activity was measured using fMRI while they watched a series of naturalistic stimuli (specifically, videos). After the scan, participants completed the UCLA loneliness scale (ULS-8) (Hays & Dimatteo, 1987). **(b)** Schematic of the analysis. We extracted time series of neural responses to the stimuli in each of 214 brain regions, and we then correlated these time series across participants to calculate inter-subject correlations (ISCs) for each dyad in each brain region.

**fMRI data acquisition.** Participants were scanned using a 3T Siemens Prisma scanner with a 32-channel head coil. Functional images were recorded using an echo-planar sequence (with echo time = 37 ms, repetition time = 800 ms, voxel size = 2.0 mm × 2.0 mm × 2.0 mm, matrix size = 104 × 104 mm, field of view = 208 mm, slice thickness = 2.0 mm, multi-band acceleration factor = 8, and 72 interleaved slices with no gap). A black screen was included at the beginning (duration = 8 seconds) and the end (duration = 20 seconds) of each run to allow the BOLD signal to stabilize. We also acquired high-resolution T1-weighted (T1w) images (with echo time = 2.48 ms, repetition time = 1,900 ms, voxel size = 1.0 mm × 1.0 mm × 1.00 mm, matrix size = 256 × 256 mm, field of view = 256 mm, slice thickness = 1.0 mm, and 208 interleaved slices with a 0.5 mm gap) for coregistration and normalization. We attached adhesive tape to the head coil in the MRI scanner and applied it across participants' foreheads; this method significantly reduces head motion (Krause et al., 2019). We used fMRIPrep version 1.4.0 for the data processing of our fMRI data (Esteban et al., 2019). See the Supplemental Material for technical details about data preprocessing.

**Cortical parcellation.** We extracted neural responses across the whole brain using the 200-parcel cortical parcellation scheme of Schaefer et al. (2018) along with 14 subcortical parcels using the Harvard–Oxford subcortical atlas (Desikan et al., 2006). This gives a total of 214 parcels.

**Inter-subject correlations (ISCs).** We extracted preprocessed time-series data and concatenated all four runs into a single time series for each participant, except for the two participants for whom we used only partial data. For these two participants, we concatenated their three usable runs into a single time series and calculated ISCs for these participants by comparing their data to the corresponding three runs in the other participants. We averaged the response time series across voxels within each of the 214 brain regions for each participant at each repetition time (TR). Our main analyses included 66 participants, so there were 2,144 unique dyads in these analyses. Our analyses that included out-degree centrality as a control variable had 63 participants, so there were a total of 1,952 unique dyads for these analyses. For each unique dyad, we calculated the Pearson correlation between the mean time series of the neural response in each of the 214 brain regions. We then took Fisher *z*-transforms of the Pearson correlations and then normalized the subsequent values (i.e., we "z-scored" them) in each brain region prior to our analyses.

**Relating ISC with loneliness.** We took the following steps to test for associations between loneliness and neural similarity in each of the 214 brain regions. First, we used a median split to stratify our sample into lonely and non-lonely groups. For conciseness, we refer to people with a ULS-8 score above the median as "lonely" and to people with a ULS-8 score at or below the median as "non-lonely". Our choice to use a median split follows the example of other recent studies that related neural similarity with behavioral measures (Finn et al., 2018; Leong et al., 2020). Whenever possible, we also conducted exploratory tests to investigate relationships between ISCs and the non-binarized loneliness variable. (We describe this in more detail below.)

To relate the dyad-level neural similarity measure with loneliness, we transformed the individual-level binarized loneliness measure into a dyad-level variable. We labeled dyads as (1) {lonely, lonely} if both individuals in a dyad were lonely, (2) {non-lonely, non-lonely} if both individuals in a dyad were non-lonely, and (3) {non-lonely, lonely} if one individual in a dyad was non-lonely and the other individual was lonely. To relate this dyad-level loneliness measure to neural similarity, we used the method of Chen et al. (2017) and fit linear mixed-effects models with crossed random effects using LME4 and LMERTEST in R (Kuznetsova et al., 2017). This approach allowed us to account for nonindependence in the data from repeated observations for each participant (i.e., because each participant is part of multiple dyads). Following the method that was suggested by Chen et al. (2017), we "doubled" the data (with redundancy) to allow fully-crossed random effects. In other words, we accounted for the symmetric nature of the ISC matrix and the fact that each participant contributes twice in a dyad (because $(i, j) = (j, i)$ for participants $i$ and $j$). See Chen et al. (2017) for more details, including information about the approach and terminology. Before performing statistical inference, we manually corrected the degrees of freedom to $N - k$, where $N$ is the number of unique observations (in our case, $N$ = 2,144) and $k$ is the number of fixed effects in the model. All findings that we report in the present paper use the corrected number of degrees of freedom. For each of our 214 brain regions, we first fit a mixed-effects model, with ISCs in the corresponding brain region as the dependent variable, the dyad-level binarized loneliness variable as the independent variable, and random intercepts for each individual (i.e., "participant 1" and "participant 2") in a dyad. We then conducted planned-contrasts using EMMEANS in R (Russell et al., 2021) to identify the brain regions where the inclusion of even one lonely individual is associated with smaller ISCs: $ISC_{\{lonely, lonely\}} > ISC_{\{non\text{-}lonely, non\text{-}lonely\}}$, $ISC_{\{lonely, lonely\}} > ISC_{\{non\text{-}lonely, lonely\}}$, and $ISC_{\{non\text{-}lonely, lonely\}} > ISC_{\{non\text{-}lonely, non\text{-}lonely\}}$. We z-scored all variables to yield standardized coefficients ($\beta$) as outputs. We FDR-corrected $p$-values for multiple comparisons at $p < 0.05$.

For our exploratory analysis in which we related a non-binarized version of the loneliness variable to ISCs, we used the maximum loneliness value of each dyad. For instance, if participant 1 in a dyad had a loneliness value of 4 and participant 2 in a dyad had a loneliness value of 6, then we assigned a loneliness value of 6 to the dyad. The choice of taking the maximum loneliness value of each dyad allowed us to test the hypothesis that only dyads with two non-lonely individuals have very similar neural responses to each other. If loneliness is associated with idiosyncratic neural responses, then the inclusion of even just one lonely individual in a dyad should be associated with low ISCs. We repeated the procedure that we described above to fit mixed-effects models with fully-crossed random effects to infer neural similarity in each brain region from the dyadic maximum loneliness. In other words, for each of the 214 brain regions, we fit a mixed-effects model, with ISCs in the corresponding brain region as the dependent variable, the maximum loneliness value of the dyad as the independent variable, and random intercepts for each individual (i.e., "participant 1" and "participant 2") in a dyad. We *z*-scored all variables, and we FDR-corrected all *p*-values for multiple comparisons at $p < 0.05$.

**Relating ISC with loneliness while controlling for overall levels of objective social disconnection, friendships between participants, and demographic similarities.** We also fit additional models to test whether or not any observed associations between loneliness and ISCs remain significant even after controlling for participants' self-reported levels of objective social disconnection, whether the individuals in a dyad were friends (because prior research suggests that friends have larger ISCs than non-friends (Hyon et al., 2020; Parkinson et al., 2018)), and dyadic similarities in all available demographic variables. (See the Supplemental Material for details about how we calculated these variables.) To do this, we fit linear mixed-effects models with fully-crossed random effects using the method of Chen et al. (2017) as described in Relating ISC with loneliness, with ISC in the corresponding brain region as the dependent variable and the dyad-level loneliness variable as the independent variable of interest, while controlling for out-degree centrality, friendships between individuals in a dyad, and dyadic similarities in age, gender, race/ethnicity, and home country by including them as covariates of no interest. We then performed planned contrasts (i.e., $ISC_{\{lonely, lonely\}} > ISC_{\{non\text{-}lonely, non\text{-}lonely\}}$, $ISC_{\{lonely, lonely\}} >$

$ISC_{\{non\text{-}lonely,\ lonely\}}$, and $ISC_{\{non\text{-}lonely,\ lonely\}} > ISC_{\{non\text{-}lonely,\ non\text{-}lonely\}}$) to test if the inclusion of one or more highly-lonely individuals in a dyad is associated with smaller ISCs while controlling for the aforementioned variables.

## Results

**Associations between binarized loneliness and neural similarity.** We first tested whether or not loneliness is associated with more idiosyncratic brain activity than one's peers. The loneliness scores in our data ranged from a minimum of 8 to a maximum of 27 (out of a possible minimum of 8 and a possible maximum of 32), with a mean of 15.91, a median of 16, a mode of 16, and a standard deviation (SD) of 4.879. Our median-split approach identified 35 individuals who had a loneliness score above the median ($n_{lonely} = 35$) and 31 individuals who had a loneliness score at or below the median ($n_{non\text{-}lonely} = 31$). (See Fig. S1 for a plot of the distribution of the loneliness scores.) To relate the dyad-level neural similarity measure with loneliness, we then transformed the individual-level binarized loneliness measure into a dyad-level variable. See Relating ISC with loneliness in the Method section for more details. Of the 2,144 unique dyads, 595 were {lonely, lonely}, 465 were {non-lonely, non-lonely}, and 1,084 were {non-lonely, lonely}.

We first used planned contrasts to compare dyadic neural similarities across the different loneliness groups and to identify brain regions where the inclusion of one or more lonely individuals in a dyad is associated with less-coordinated neural responses (i.e., $ISC_{\{lonely,\ lonely\}} > ISC_{\{non\text{-}lonely,\ non\text{-}lonely\}}$, $ISC_{\{lonely,\ lonely\}} > ISC_{\{non\text{-}lonely,\ lonely\}}$, and $ISC_{\{non\text{-}lonely,\ lonely\}} > ISC_{\{non\text{-}lonely,\ non\text{-}lonely\}}$). We found that dyads in which both individuals were lonely had smaller ISCs than dyads in which both members were non-lonely (see Fig. 2 and Tables S2–S4) in the dorsal medial prefrontal cortex (DMPFC), ventrolateral prefrontal cortex (VLPFC), dorsal lateral prefrontal cortex (DLPFC), precuneus, posterior cingulate cortex, superior temporal cortex (including the superior temporal sulcus (STS)), superior parietal lobule (SPL), and inferior parietal lobule (IPL). The ISCs in several subcortical regions — specifically, the left nucleus accumbens, left caudate nucleus, and right pallidum — were also smaller in {lonely, lonely} dyads than in {non-lonely, non-lonely} dyads (see Fig. 2 and Tables S5–S6).

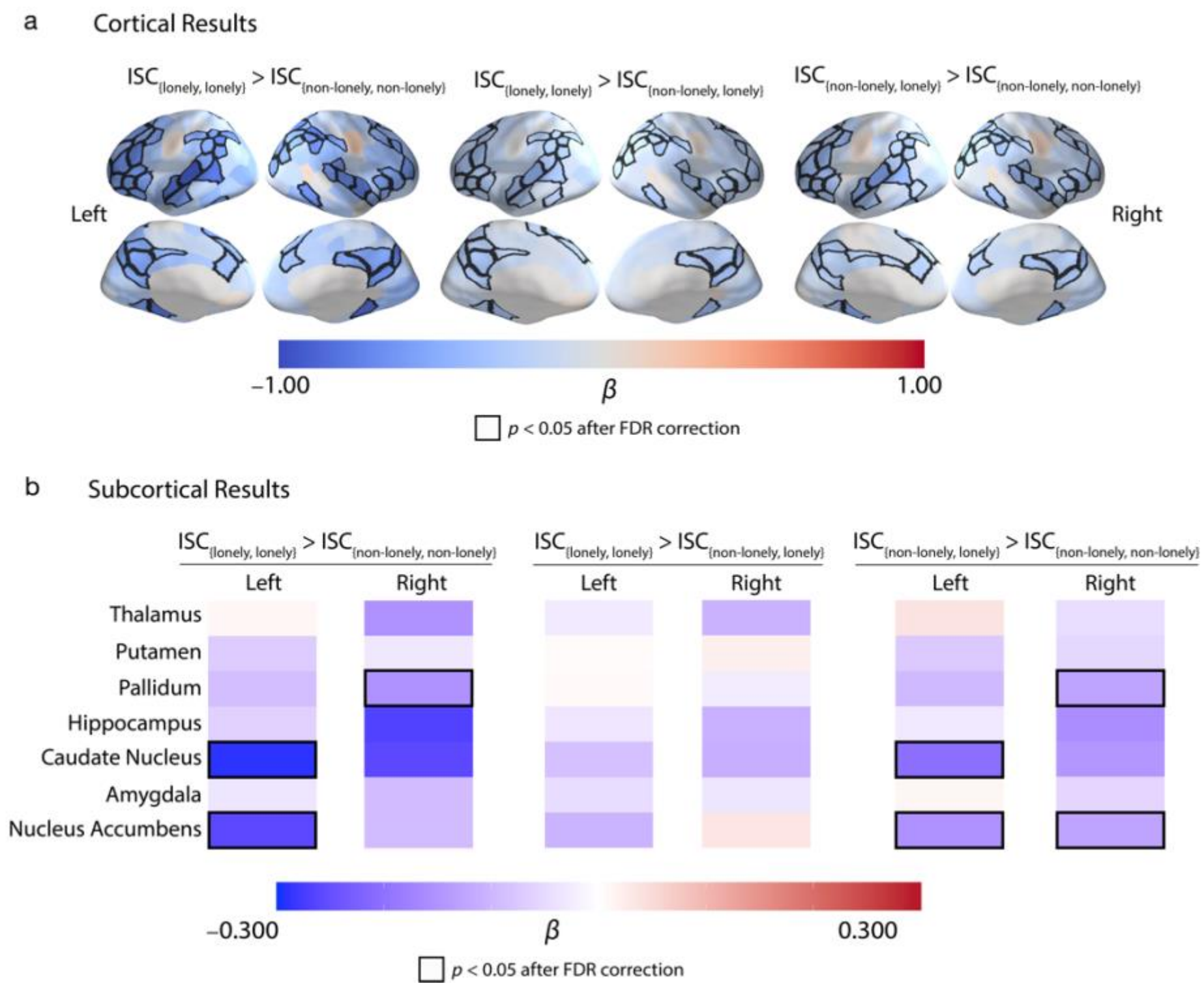

**Fig. 2.** Linking greater levels of loneliness to more-idiosyncratic neural responses. **(a)** The ISCs were smaller in brain regions such as the DMPFC, precuneus, VLPFC, DLPFC, STS, IPL, and SPL in dyads in which both individuals were lonely (i.e., {lonely, lonely}) than in dyads in which neither individual was lonely (i.e., {non-lonely, non-lonely}). We observed similar patterns when we compared dyads in which both individuals were lonely (i.e., {lonely, lonely}) to dyads with one non-lonely individual and one lonely individual (i.e., {non-lonely, lonely}) and when we compared dyads with one non-lonely individual and one lonely individual (i.e., {non-lonely, lonely}) to dyads with two non-lonely individuals (i.e., {lonely, lonely}). **(b)** The ISCs were smaller in the right pallidum, left caudate nucleus, and left nucleus accumbens in dyads with two lonely individuals than in dyads with two non-lonely individuals. The labels "Left" and "Right" refer to the hemispheres of the brain regions that are listed in the left panel. The quantity $\beta$ is the standardized regression coefficient. Regions with significant associations between loneliness and ISC are outlined in black (using an FDR-corrected significance threshold of $p < 0.05$).

**Associations between non-binarized loneliness and neural similarity.** We also related mean ISCs to a non-binarized loneliness value (specifically, maximum loneliness value in the dyads, as we discussed in Relating ISC with loneliness in the Method section). As with our results with the binarized loneliness variable, we found a negative association between maximum loneliness and neural similarity in the VLPFC, DLPFC, precuneus, posterior cingulate cortex, superior temporal cortex, IPL, and SPL. In other words, mirroring our results for the binarized loneliness measure, dyads in which one or both individuals were lonely (as indicated by a larger maximum loneliness) were characterized by less neural similarity than dyads in which both individuals were non-lonely (as indicated by a smaller maximum loneliness) (see Fig. S2).

In Fig. 3, we show the relationship between loneliness and ISC in a parcel in the left IPL, in which neural similarity has been associated previously both with shared understanding of events (Nguyen et al., 2019; Yeshurun et al., 2017) and with similarities in perspectives (Lahnakoski et al., 2014). This figure shows the participant-by-participant ISC matrix for this parcel, with rows and columns ordered based on participants' loneliness scores. The ISCs tend to become smaller as one moves to the right and down along the diagonal, indicating that the smallest ISCs tend to occur in increasingly lonely dyads.

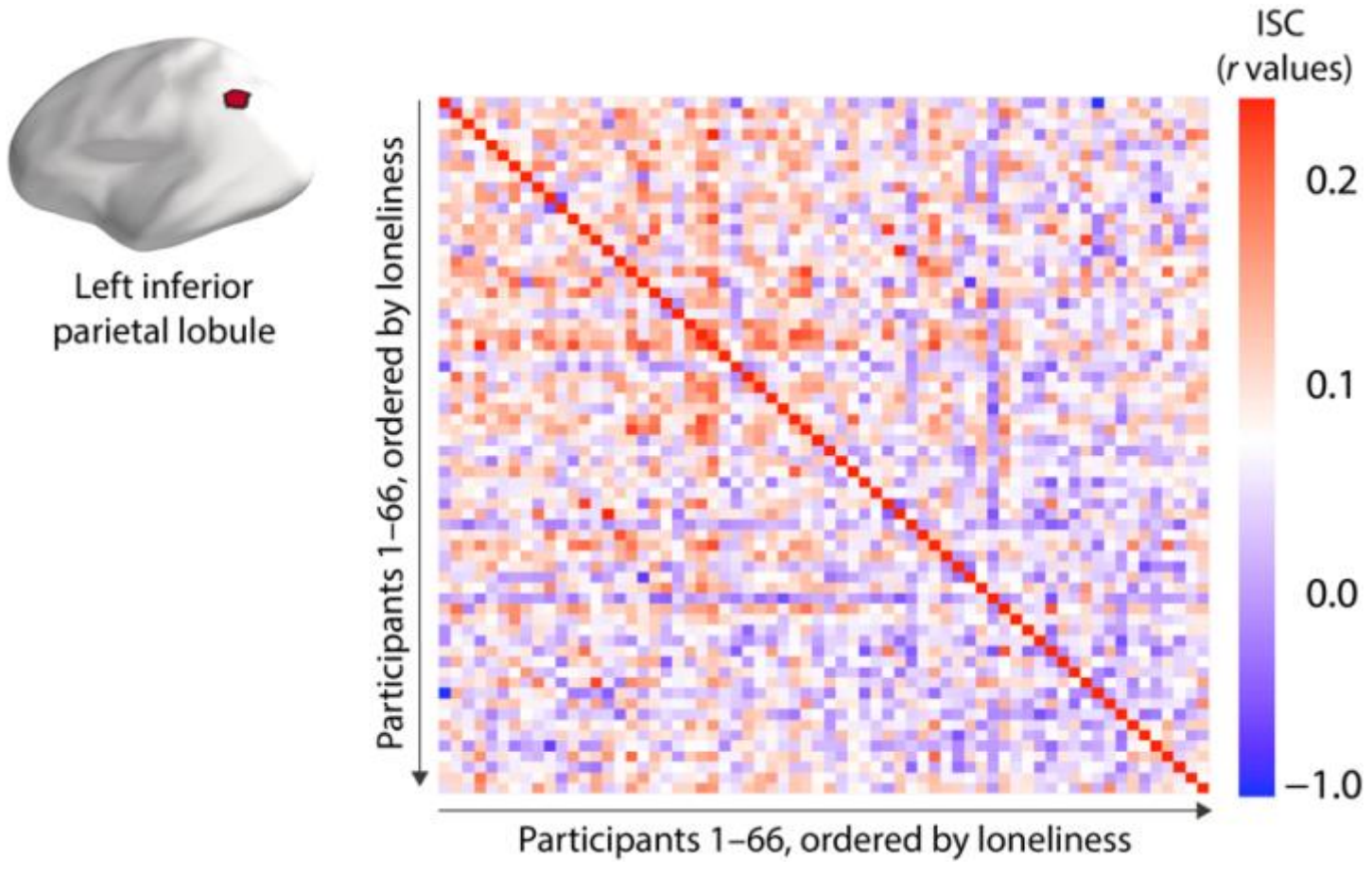

**Fig. 3.** Associations between ISCs and loneliness in the left IPL. The ISC matrix for a parcel in the left IPL; the rows and columns of the matrix are ordered by increasing loneliness. The largest ISC values tend to occur in the top-left corner of the matrix (as indicated by warm colors). This corner has the dyads with the lowest loneliness scores. The bottom and right portions of the matrix tend to have the smallest ISC values (as indicated by cool colors). The pattern of results in this matrix supports the hypothesis that having at least one lonely person in a dyad is associated with smaller ISCs.

**Associations between loneliness and neural similarity when controlling for objective social disconnection, friendships between participants, and demographic similarities.** We also tested if the associations between loneliness and ISCs remain significant even after controlling for participants' self-reported levels of objective social disconnection, whether or not the individuals in a dyad were friends, and dyadic similarities in all available demographic variables. As in our main results, we found that ISCs in the VLPFC, DLPFC, superior temporal cortex, SPL, and IPL were smaller in dyads in which both individuals were lonely ({lonely, lonely}) than in dyads in which both individuals were not lonely ({non-lonely, non-lonely}) (see Fig. 4a and Table S7). Additionally, ISCs in the left nucleus accumbens were smaller in {lonely, lonely} dyads than in {non-lonely, non-lonely} dyads (see Fig. 4b and Table S8). We observed similar patterns when we compared {lonely, lonely} dyads to {non-lonely, lonely} dyads and {non-lonely, lonely} dyads to {non-lonely, non-lonely} dyads (see Fig. 4b and Tables S9–S11).

Given recent findings that link neural similarity with popularity (Baek et al., 2022), we also fit models that control for participants' in-degree centrality (i.e., the number of times that the participant was nominated as a friend by others in their social network). The results of these analyses (see Fig. S3) are similar to those of our main analyses.

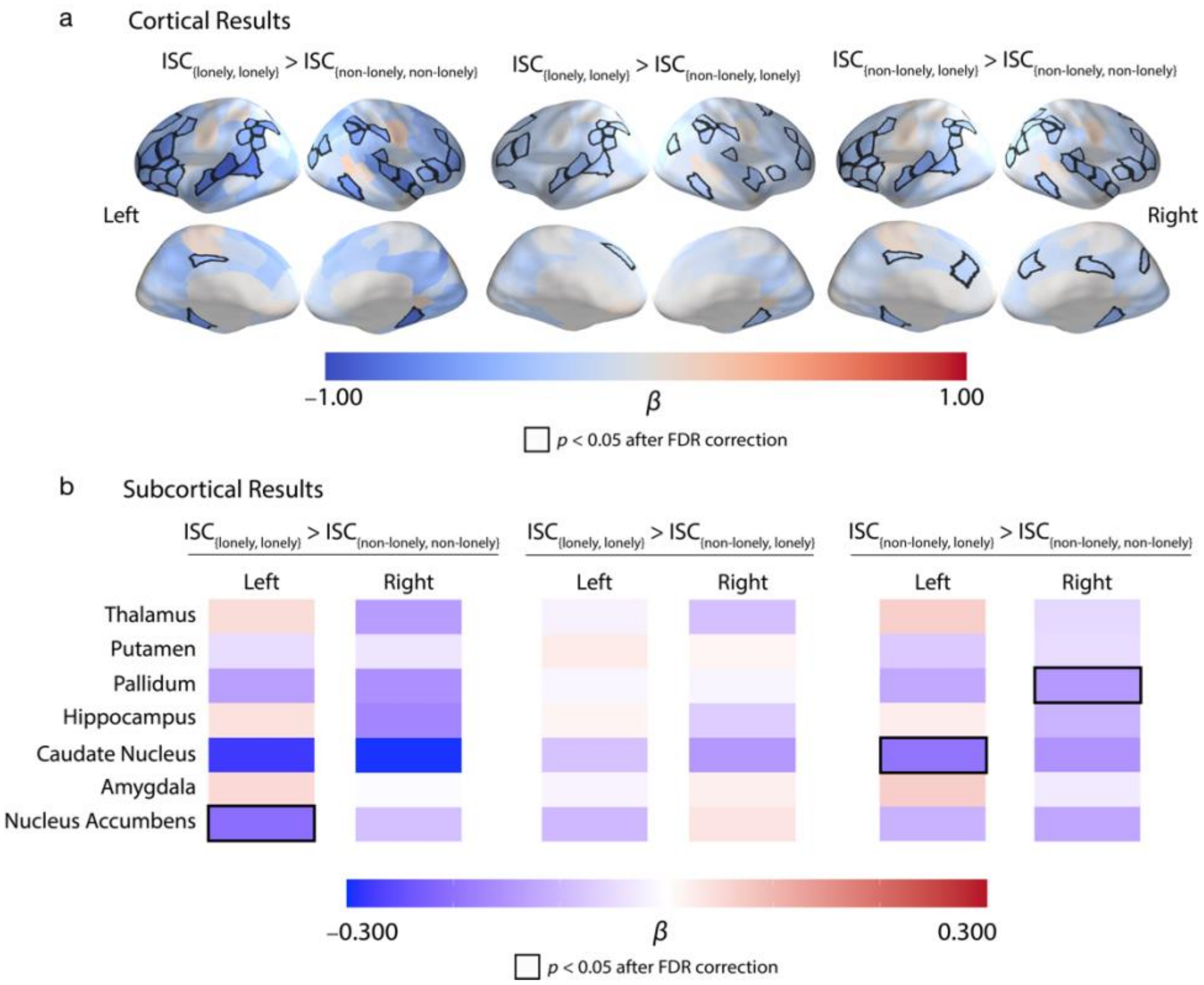

**Fig. 4.** Linking loneliness to idiosyncratic neural responses while controlling for objective social disconnection, demographic similarities, and friendships between participants. **(a)** As in our main results, ISCs were smaller in brain regions (including the VLPFC, DLPFC, STS, IPL, and SPL) that are associated with social cognition, shared understanding of events, and friendship in dyads with individuals who were both lonely (i.e., {lonely, lonely}) than in dyads in which both individuals were non-lonely (i.e., {non-lonely, non-lonely}). We observed similar patterns when we compared dyads with individuals who were both lonely (i.e., {lonely, lonely}) to dyads with one non-lonely individual and one lonely individual (i.e., {non-lonely, lonely}) and when we compared dyads with one non-lonely individual and one lonely individual (i.e., {non-lonely, lonely}) to dyads with two non-lonely individuals (i.e., {non-lonely, non-lonely}). **(b)** The ISCs were smaller in the left nucleus accumbens in dyads with two lonely individuals than in dyads with two non-lonely individuals. The labels "Left" and "Right" refer to the hemispheres of the brain regions that are listed in the left panel. The quantity $\beta$ is the standardized regression coefficient. Regions with significant associations between loneliness and ISC are outlined in black (using an FDR-corrected significance threshold of $p < 0.05$).

## Discussion

Our results suggest that lonely people process the world idiosyncratically, which may contribute to the reduced sense of being understood that often accompanies loneliness. In other words, we found that non-lonely individuals were very similar to each other in their neural responses, whereas lonely individuals were remarkably dissimilar to each other and to their non-lonely peers. Our results remained significant even after controlling for individuals' objective levels of social connection/disconnection (specifically, their numbers of friends), demographic variables, and friendships between participants. Therefore, we conclude that lonely people may view the world in a way that is different from their peers. These findings suggest the possibility that being surrounded predominantly by people who view the world differently from oneself may be a risk factor for loneliness (even if one socializes regularly with them).

Brain areas in which responses were more similar include the DMPFC, precuneus, VLPFC, DLPFC, STS, SPL, IPL, and nucleus accumbens. Notably, in many of these regions — in particular, those that belong to the default-mode network — similar neural responding when watching videos has been associated with similarities in understanding and interpretation of narratives (Finn et al., 2018; Nguyen et al., 2019; Yeshurun et al., 2017) and friendship (Parkinson et al., 2018). We found that this pattern remained similar even after controlling for objective social connection/disconnection, demographic similarities, and friendships between participants. Consequently, our results suggest that lonely individuals process the world in a way that is dissimilar from their peers as well as each other. Future work can further test this possibility by using behavioral experimentation and semantic analyses to examine what aspects of lonely individuals' interpretations are particularly idiosyncratic.

Although it is unclear whether the observed idiosyncratic processing in lonely individuals is a cause or a result of loneliness, the associated lack of shared understanding may lead to challenges at achieving social connections. These effects hold even after controlling for the number of friends of individuals and whether or not two individuals in a dyad are friends with each other, suggesting that our findings are not merely a consequence of lonely people being less likely to have friends or non-lonely individuals being more likely to be friends with each other. Instead, we observed that individuals with high levels of loneliness — regardless of the number of their objective social connections — were more likely to have idiosyncratic neural responses. It is also likely that the extent of objective and subjective social connection/disconnection fluctuates with time, which may in turn influence or be influenced by the extent to which an individual processes the world idiosyncratically. Future work that implements a longitudinal design may further elucidate the causal directions of these relationships.

In our study, we collected the data for characterizing subjective and objective social connection/disconnection during two different time periods. We obtained the fMRI data and the loneliness measures between September and November, whereas we collected the social-network survey data in December and January. Although the time gap between the two data collections was small, future work that tests the relationships between neural similarity, subjective social connection/disconnection, and objective social connection/disconnection measures that are obtained simultaneously and during multiple time periods can enrich our understanding of how these variables interact and change with time.

Our findings dovetail with recent suggestions that the default-mode network is a 'sense-making' network that combines extrinsic information about an individual's environment with existing internal information of their past memories and knowledge (Yeshurun et al., 2021) and findings that suggest that loneliness is associated with structural and functional differences in the default-mode network (Spreng et al., 2020). For instance, loneliness has been associated with greater variation in the gray-matter volume in the default-mode network, suggesting that there is greater idiosyncrasy in the *structure* of the default-mode network in lonely individuals than in non-lonely individuals. We found that lonely individuals have idiosyncratic *functional* brain responses to audiovisual stimuli in regions of the default-mode network, whereas non-lonely individuals were exceptionally similar to each other. Our findings thereby add further insight into the idiosyncrasies of the default-mode network that may characterize lonely individuals.

What types of idiosyncratic thought processes characterize lonely individuals? Prior research offers some clues about what thought processes may potentially lead to idiosyncrasies in neural responses in lonely individuals. In our study, lonely individuals had smaller ISCs in subcortical regions (such as the left nucleus accumbens) that constitute part of the brain's reward system (Knutson et al., 2001) and in regions of the lateral posterior parietal cortex that are associated with bottom-up and top-down orienting of attention (Corbetta & Shulman, 2002). Therefore, one possibility is that lonely individuals do not find value in the same aspects of situations or scenes as their peers (and instead focus on other aspects of situations in an idiosyncratic fashion), perhaps due to differences in their preferences, expectations, and/or memories that can in turn shape how they attend to and interpret stimuli. This may result in a reinforcing feedback loop in which lonely individuals perceive themselves to be different from their peers, which may in turn lead to further challenges in achieving social connection. Indeed, in one recent study (Courtney & Meyer, 2020), greater loneliness was associated with reduced neural representational similarity between oneself and others in the medial prefrontal cortex, suggesting that lonely individuals think of themselves in a way that is more dissimilar to others than is the case for non-lonely individuals. Exceptional dissimilarities between lonely individuals and their peers in how they process the world may contribute to an overall sense of lacking shared understanding that often accompanies loneliness.

In summary, our findings suggest that processing the world differently from those around oneself is linked to loneliness. These findings were reflected in lonely individuals' idiosyncratic neural responses in brain regions that have been associated with shared interpretations of events, attentional orienting, and reward processing. Moreover, these effects remained significant even after controlling for objective social disconnection and friendships between individuals. Therefore, being surrounded by people who view the world differently from oneself may be a risk factor for loneliness, even if one socializes regularly with them.

## Open Practices Statement

The data, code, and materials for the study will be made available upon request to the corresponding author. The study was not preregistered.

Supplemental Materials for "**Lonely individuals process the world in idiosyncratic ways**"

Elisa C. Baek*[1], Ryan Hyon[1], Karina López[1], Meng Du[1], Mason A. Porter[2,3], and Carolyn Parkinson*[1,4]

[1]Department of Psychology, University of California, Los Angeles, [2] Department of Mathematics, University of California, Los Angeles, [3]Sante Fe Institute, [4] Brain Research Institute, University of California, Los Angeles

* Corresponding authors

## Supplementary method

**fMRI data analysis.** We used fMRIPrep version 1.4.0 to process our fMRI data (Esteban et al., 2019). We have taken the descriptions of anatomical and functional data preprocessing in this section from the recommended boilerplate text that is generated by fMRIPrep and released under a CC0 license, with the intention that researchers reuse the text to facilitate clear and consistent descriptions of preprocessing steps (and thereby enhance the reproducibility of studies).

For each participant, the T1-weighted (T1w) image was corrected for intensity non-uniformity (INU) with N4BiasFieldCorrection, distributed with ANTs 2.1.0 (Avants et al., 2011), and used as T1w-reference throughout the workflow. Brain tissue segmentation of cerebrospinal fluid (CSF), white matter (WM), and gray matter (GM) was performed on the brain-extracted T1w using FSL fast (Smith et al., 2004). Volume-based spatial normalization to the ICBM 152 Nonlinear Asymmetrical template version 2009c (MNI152NLin2009cAsym) was performed through nonlinear registration with antsRegistration (ANTs 2.1.0; Avants et al., 2011).

For each of the BOLD runs of each participant, the following preprocessing was performed. First, a reference volume and its skull-stripped version were generated using a custom methodology of fMRIPrep. The BOLD reference was then coregistered to the T1w reference using FSL flirt (Smith et al., 2004) with the boundary-based registration cost function. Coregistration was configured with nine degrees of freedom to account for distortions remaining in the BOLD reference. Head-motion parameters with respect to the BOLD reference (transformation matrices, and six corresponding rotation and translation parameters) were estimated before any spatiotemporal filtering using FSL mcflirt (Smith et al., 2004). Automatic removal of motion artifacts using independent component analysis (ICA-AROMA) was performed on the preprocessed BOLD on MNI-space time series after removal of non-steady-state volumes and spatial smoothing with an isotropic, Gaussian kernel of 6mm FWHM (full-width half-maximum). The BOLD time series were then resampled to the MNI152Nlin2009cAsym standard space.

The following 10 confounding variables generated by fMRIPrep were included as nuisance regressors: global signals extracted from within the cerebrospinal fluid, white matter, and whole-brain masks; framewise displacement; three translational motion parameters; and three rotational motion parameters.

**Calculation of control variables.** We took an analogous approach to the one that we described in Relating ISC with loneliness in the Method section of the main manuscript to transform the individual-level out-degree centrality variable into a dyad-level variable to control for objective social disconnection. First, using each participant's response to the social-network survey, we calculated their out-degree centrality (i.e., the number of people in their residential community that the participant nominated as a friend). As in our approach in Relating ISC with loneliness in the Method section of the main manuscript, we used a median split to label each individual as having a "high" or "low" level of objective social disconnection. We categorized participants into the low objective social-disconnection group if they had an out-degree that was larger than the median and into the high objective social-disconnection group if they had an out-degree that was less than or equal to the median. We then transformed the individual-level out-degree centrality variable into a dyad-level variable using the method that we described in Relating ISC with loneliness in the Method section of the main manuscript to transform the individual-level out-degree variable into a dyad-level variable.

To control for similarities in age, for each dyad, we computed the absolute value of the difference between the ages of the two individuals in the dyad (i.e., age_difference = $|age_{participant_1} - age_{participant_2}|$). We then transformed this difference into a similarity measure, such that larger numbers indicate greater similarities. Specifically, we calculated age_similarity = 1 – (age_difference/max(age_difference). To control for similarities in self-reported gender, we created an indicator variable in which 0 signifies different genders and 1 signifies the same gender. To control for similarities in ethnicity, we created an indicator variable for each race/ethnicity category (Asian, Black/African, Hispanic/Latinx, Native American, Pacific Islander, and Caucasian/White) in which 0 signifies a different self-reported race/ethnicity and 1 signifies the same self-reported race/ethnicity. The participants were able to self-report as many races/ethnicities as they desired. For each dyad, we created an overall indicator variable for race/ethnicity in which 0 signifies no shared race/ethnicity and 1 signifies at least one shared race/ethnicity. That is, if the two individuals in a dyad self-reported even one common race/ethnicity, we coded them as having a shared race/ethnicity. To control for similarities in home country, we created an additional indicator variable in which 0 signifies different home countries and 1 signifies the same home country. If either individual in a dyad nominated the other as a friend in the social-network survey, we coded the dyad as signifying an undirected friendship.

## Supplementary table: Descriptions of stimuli

Table S1. Descriptions of stimuli

| | Run # | Video | Content |
|---|---|---|---|
| 1 | 1 | An Astronaut's View of Earth | An astronaut discusses viewing Earth from space and, in particular, witnessing the effects of climate change from space. He then urges viewers to mobilize to address this issue. |
| 2 | 1 | All I Want | A sentimental music video depicting a social outcast with a facial deformity who is seeking companionship. |
| 3 | 1 | Scientific demonstration | An astronaut at the International Space Station demonstrates and explains what happens when one wrings out a waterlogged washcloth in space. |
| 4 | 1 | Food Inc. | An excerpt from a documentary discussing how the fast-food industry influences food production and farming practices in the United States. |
| 5 | 2 | We Can Be Heroes | An excerpt from a mockumentary-style series in which a man discusses why he nominated himself for the title of Australian of the Year. |
| 6 | 2 | Ban College Football | Journalists and athletes debate whether football should be banned as a college sport. |
| 7 | 2 | Soccer match | Highlights from a soccer match. |
| 8 | 2 | Ew! | A comedy skit in which grown men play teenage girls disgusted by the things around them. |
| 9 | 2 | Life's Too Short | An example of a 'cringe comedy' in which a dramatic actor is depicted unsuccessfully trying his hand at improvisational comedy. |
| 10 | 2 | America's Funniest Home Videos | A series of homemade video clips that depict examples of unintentional physical comedy arising from accidents. |
| 11 | 3 | Zima Blue | A philosophical, animated short set in a futuristic world. |
| 12 | 3 | Nathan For You | An episode from a 'docu-reality' comedy in which the host convinces people, who are not always in on the joke, to engage in a variety of strange behaviors. |
| 13 | 4 | College Party | An excerpt from a film depicting a party scene in which a bashful college student is pressured to drink alcohol. |
| 14 | 4 | Eighth Grade | Two excerpts from a film. They depict (1) a young teenager who video blogs about her mental-health issues and (2) an awkward scene between two teenagers on a dinner date. |

Note: The video files for the stimuli that we used in this study are available at https://gitlab.com/anon_authors/dorm_study_stimuli

## Supplementary figure: Distribution of loneliness

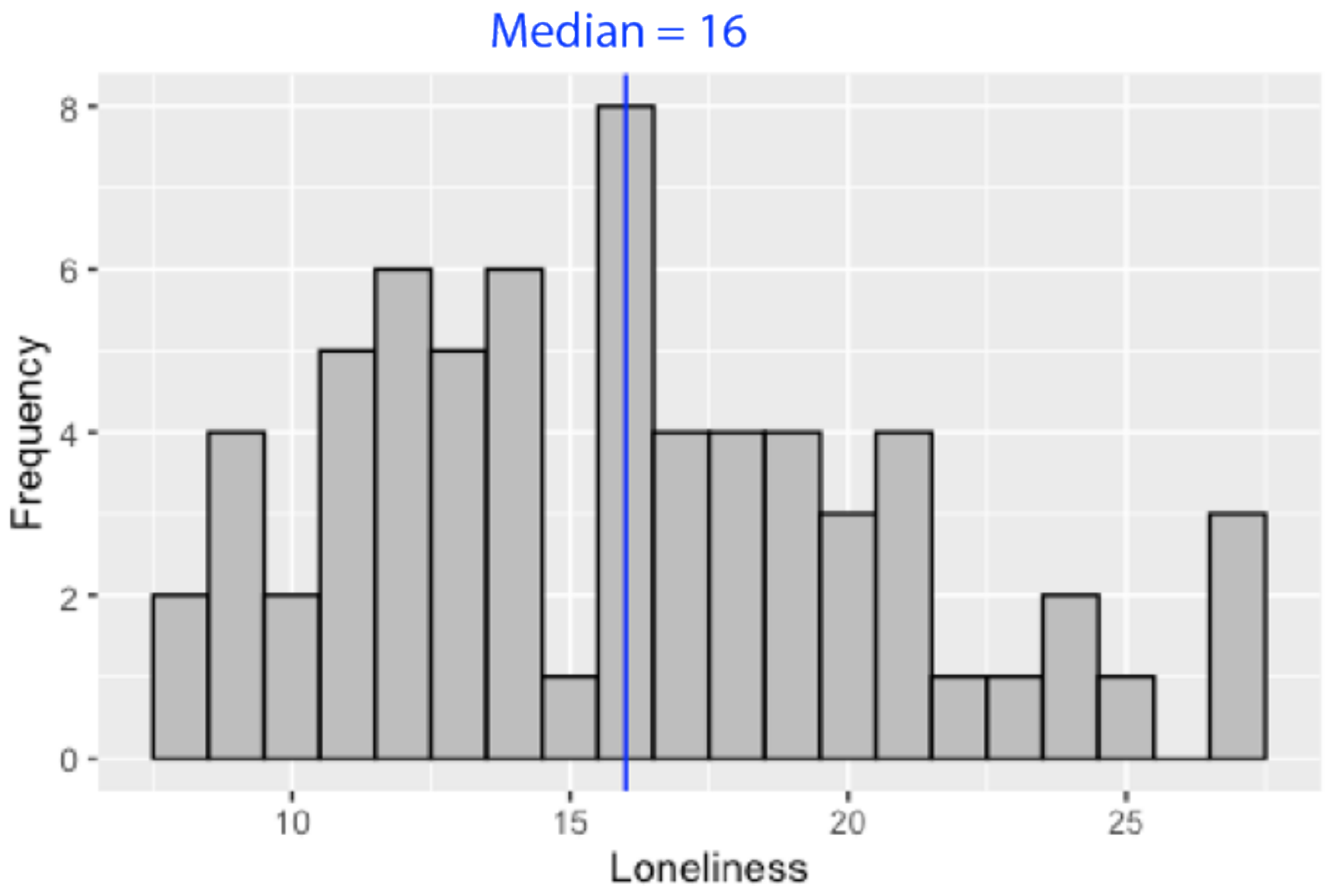

**Fig. S1.** Distribution of loneliness scores of the present study's participants. We measured loneliness using the UCLA Loneliness Scale (ULS-8) *(36)*.

**Supplementary figure: Results associating neural similarity with maximum loneliness in dyads**

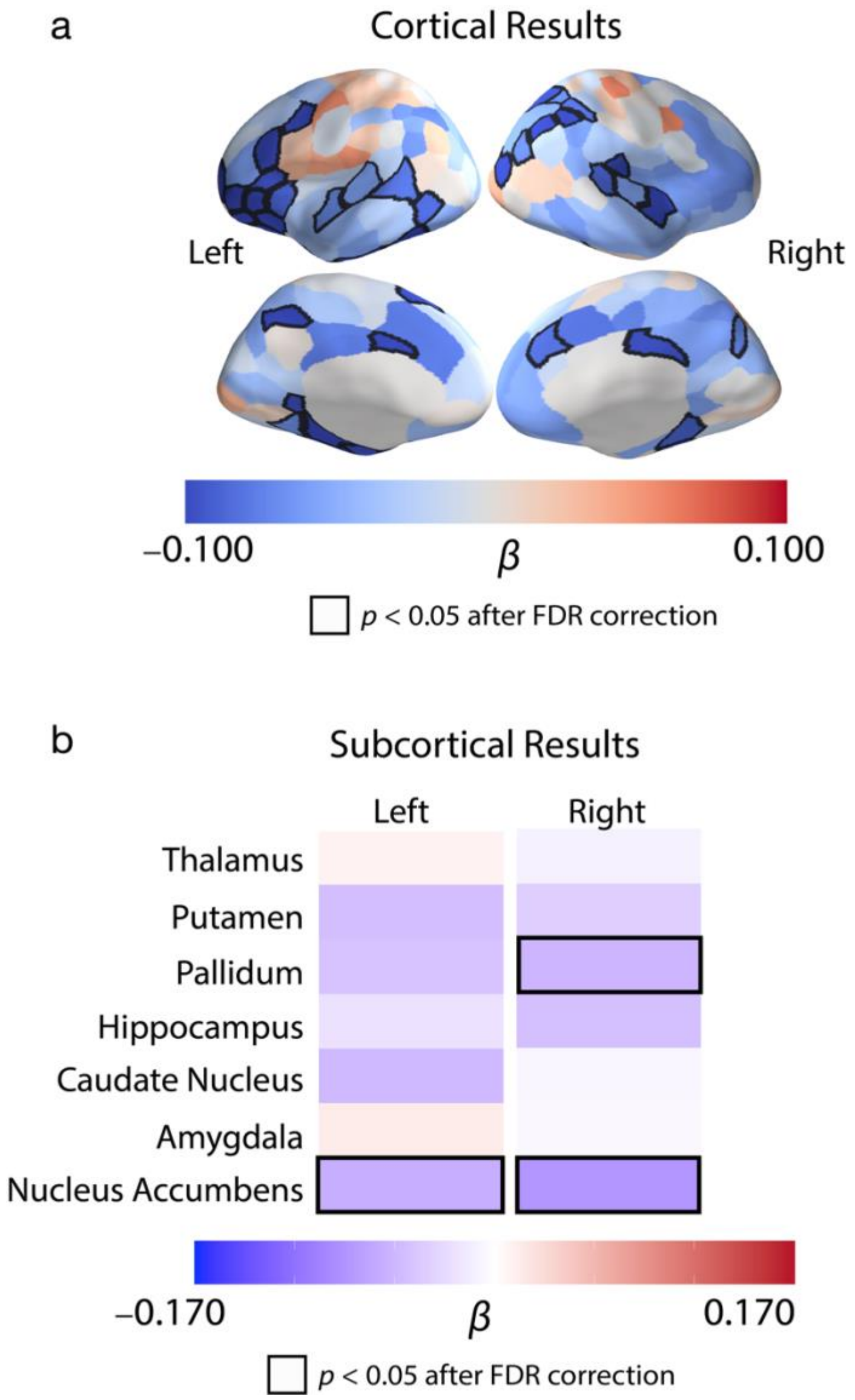

**Fig. S2.** Associating neural similarity with the maximum loneliness values in dyads. **(a)** We observed a negative association between ISCs and the maximum loneliness value. ISCs in brain regions (including the VLPFC, DLPFC, precuneus, posterior cingulate cortex, superior temporal cortex, IPL, and SPL) that are associated with social cognition and shared understanding of events were associated with lower values of maximum loneliness. **(b)** There was a negative association between ISCs and maximum loneliness in the left and right nucleus accumbens and the right pallidum. The labels "Left" and "Right" refer to the hemispheres of the brain regions that are listed in the left panel. The quantity $\beta$ is the standardized regression coefficient. Regions with significant associations between the maximum loneliness and ISC are outlined in black (using an FDR-corrected significance threshold of $p < 0.05$).

## Supplementary results: Supplementary tables

In this section, we give tables of numerical values that are associated with various figures in the main manuscript.

Table S2. Results linking loneliness and neural responses: Cortical results (corresponding to Fig. 2a of the main manuscript)

Contrast: $ISC_{\{lonely, lonely\}} > ISC_{\{non\text{-}lonely, non\text{-}lonely\}}$

| Hemisphere | Network | Component Name[1] | Parcel Number | $\beta$ | SE | $p$-value (corrected) |
|---|---|---|---|---|---|---|
| Left | Control | Parietal | 1 | −0.736 | 0.209 | 0.000 |
| Right | Dorsal Attention | Posterior | 4 | −0.783 | 0.250 | 0.001 |
| Right | Salience / Ventral Attention | Frontal Operculum | 2 | −0.713 | 0.232 | 0.001 |
| Left | Default | Parahippocampal Cortex | 1 | −0.758 | 0.249 | 0.001 |
| Right | Control | Parietal | 1 | −0.649 | 0.219 | 0.001 |
| Right | Control | Parietal | 2 | −0.675 | 0.228 | 0.001 |
| Right | Control | Lateral Prefrontal Cortex | 3 | −0.598 | 0.204 | 0.001 |
| Left | Somatomotor | Somatomotor | 2 | −0.897 | 0.308 | 0.001 |
| Right | Control | Lateral Prefrontal Cortex | 6 | −0.547 | 0.190 | 0.001 |
| Right | Control | Temporal | 1 | −0.678 | 0.244 | 0.002 |
| Right | Visual | Visual | 2 | −0.786 | 0.287 | 0.002 |
| Left | Dorsal Attention | Posterior | 7 | −0.774 | 0.283 | 0.002 |
| Right | Control | Precuneus | 1 | −0.713 | 0.263 | 0.002 |
| Left | Control | Lateral Prefrontal Cortex | 1 | −0.575 | 0.214 | 0.002 |
| Left | Default | Temporal | 5 | −0.792 | 0.298 | 0.003 |
| Right | Somatomotor | Somatomotor | 4 | −0.582 | 0.223 | 0.003 |
| Left | Control | Parietal | 2 | −0.643 | 0.249 | 0.003 |
| Left | Control | Lateral Prefrontal Cortex | 3 | −0.470 | 0.183 | 0.003 |
| Left | Default | Prefrontal Cortex | 12 | −0.520 | 0.203 | 0.003 |
| Left | Default | Temporal | 8 | −0.620 | 0.241 | 0.003 |
| Left | Somatomotor | Somatomotor | 1 | −0.768 | 0.300 | 0.003 |
| Left | Salience / Ventral Attention | Frontal Operculum | 2 | −0.534 | 0.209 | 0.003 |
| Left | Default | Prefrontal Cortex | 1 | −0.537 | 0.211 | 0.003 |
| Left | Control | Parietal | 3 | −0.640 | 0.259 | 0.005 |
| Left | Control | Lateral Prefrontal Cortex | 5 | −0.660 | 0.270 | 0.005 |
| Right | Dorsal Attention | Posterior | 8 | −0.692 | 0.283 | 0.005 |
| Right | Salience / Ventral Attention | Frontal Operculum | 1 | −0.459 | 0.188 | 0.005 |
| Left | Salience / Ventral Attention | Parietal Operculum | 3 | −0.539 | 0.222 | 0.005 |
| Left | Control | Lateral Prefrontal Cortex | 2 | −0.510 | 0.211 | 0.005 |
| Right | Salience / Ventral Attention | Frontal Operculum | 3 | −0.534 | 0.222 | 0.006 |
| Left | Limbic | Orbital Frontal Cortex | 1 | −0.405 | 0.170 | 0.006 |
| Left | Control | Lateral Prefrontal Cortex | 4 | −0.618 | 0.259 | 0.006 |
| Right | Visual | Visual | 11 | −0.624 | 0.264 | 0.006 |
| Left | Default | Posterior Cingulate Cortex | 3 | −0.452 | 0.195 | 0.007 |
| Right | Default | Posterior Cingulate Cortex | 1 | −0.628 | 0.274 | 0.008 |
| Right | Control | Medial Posterior Prefrontal Cortex | 1 | −0.525 | 0.230 | 0.008 |
| Left | Default | Prefrontal Cortex | 13 | −0.553 | 0.244 | 0.009 |
| Left | Default | Prefrontal Cortex | 10 | −0.470 | 0.213 | 0.011 |
| Right | Default | Parietal | 1 | −0.564 | 0.256 | 0.011 |
| Left | Control | Lateral Prefrontal Cortex | 6 | −0.535 | 0.244 | 0.011 |
| Left | Default | Prefrontal Cortex | 5 | −0.592 | 0.273 | 0.011 |
| Right | Dorsal Attention | Posterior | 3 | −0.539 | 0.250 | 0.012 |
| Right | Somatomotor | Somatomotor | 1 | −0.664 | 0.311 | 0.013 |
| Left | Control | Precuneus | 1 | −0.539 | 0.253 | 0.013 |
| Left | Default | Prefrontal Cortex | 3 | −0.523 | 0.246 | 0.013 |
| Left | Default | Prefrontal Cortex | 8 | −0.384 | 0.181 | 0.013 |
| Right | Default | Posterior Cingulate Cortex | 2 | −0.546 | 0.257 | 0.013 |
| Right | Default | Ventral Prefrontal Cortex | 1 | −0.509 | 0.241 | 0.014 |
| Right | Control | Medial Posterior Prefrontal Cortex | 3 | −0.379 | 0.181 | 0.015 |
| Right | Control | Ventral Prefrontal Cortex | 1 | −0.449 | 0.215 | 0.015 |
| Right | Visual | Visual | 15 | −0.634 | 0.303 | 0.015 |
| Left | Salience / Ventral Attention | Parietal Operculum | 1 | −0.548 | 0.267 | 0.016 |
| Right | Default | Medial Prefrontal Cortex | 7 | −0.439 | 0.216 | 0.017 |
| Right | Control | Parietal | 3 | −0.501 | 0.248 | 0.018 |
| Right | Default | Parietal | 3 | −0.486 | 0.241 | 0.018 |
| Left | Default | Prefrontal Cortex | 11 | −0.433 | 0.216 | 0.018 |
| Left | Default | Posterior Cingulate Cortex | 2 | −0.524 | 0.262 | 0.019 |
| Left | Default | Posterior Cingulate Cortex | 1 | −0.515 | 0.259 | 0.019 |
| Left | Default | Posterior Cingulate Cortex | 4 | −0.510 | 0.258 | 0.020 |
| Right | Dorsal Attention | Frontal Eye Fields | 2 | −0.481 | 0.244 | 0.020 |
| Left | Default | Prefrontal Cortex | 9 | −0.477 | 0.243 | 0.021 |
| Right | Dorsal Attention | Posterior | 10 | −0.490 | 0.251 | 0.022 |
| Left | Default | Temporal | 3 | −0.584 | 0.301 | 0.023 |
| Right | Somatomotor | Somatomotor | 2 | −0.595 | 0.310 | 0.024 |
| Right | Visual | Visual | 1 | −0.513 | 0.273 | 0.028 |
| Left | Dorsal Attention | Posterior | 10 | −0.504 | 0.271 | 0.029 |
| Left | Salience / Ventral Attention | Lateral Prefrontal Cortex | 1 | −0.377 | 0.204 | 0.031 |
| Left | Visual | Visual | 1 | −0.545 | 0.297 | 0.032 |
| Left | Dorsal Attention | Frontal Eye Fields | 2 | −0.476 | 0.261 | 0.032 |
| Right | Dorsal Attention | Posterior | 7 | −0.504 | 0.279 | 0.035 |
| Right | Control | Lateral Prefrontal Cortex | 7 | −0.423 | 0.235 | 0.035 |
| Left | Dorsal Attention | Posterior | 4 | −0.483 | 0.279 | 0.043 |
| Right | Control | Medial Posterior Prefrontal Cortex | 4 | −0.423 | 0.245 | 0.044 |
| Left | Dorsal Attention | Posterior | 1 | −0.455 | 0.269 | 0.048 |

[1]Component names reflect the labels that were provided by Yeo et al. (2011). These labels split the seven networks into spatially connected components. We provide the component names and intra-component parcel numbers to aid in the identification of the parcels.

Note: The component names and the parcel numbers (within components) correspond to those that accompany the Schaefer et al., (2018) parcellation. Each row represents a single parcel in the brain. Component names reflect the component labels. More information about each parcel is available at their GitHub page (https://github.com/ThomasYeoLab/CBIG/tree/master/stable_projects/brain_parcellation/Schaefer2018_LocalGlobal). The quantity $\beta$ is the standardized regression coefficient and the quantity SE is the standard error.

Table S3. Results linking loneliness and neural responses: Cortical results (corresponding to Fig. 2a of the main manuscript)

Contrast: $ISC_{\{lonely, lonely\}} > ISC_{\{non-lonely, lonely\}}$

| Hemisphere | Network | Component Name[1] | Parcel Number | *β* | SE | *p*-value (corrected) |
|---|---|---|---|---|---|---|
| Left | Control | Parietal | 1 | −0.343 | 0.109 | 0.001 |
| Right | Dorsal Attention | Posterior | 4 | −0.370 | 0.128 | 0.001 |
| Right | Salience / Ventral Attention | Frontal Operculum | 2 | −0.335 | 0.120 | 0.002 |
| Right | Control | Lateral Prefrontal Cortex | 6 | −0.281 | 0.101 | 0.002 |
| Left | Dorsal Attention | Posterior | 7 | −0.391 | 0.143 | 0.002 |
| Right | Control | Lateral Prefrontal Cortex | 3 | −0.286 | 0.107 | 0.002 |
| Left | Default | Prefrontal Cortex | 12 | −0.282 | 0.107 | 0.003 |
| Left | Salience / Ventral Attention | Parietal Operculum | 3 | −0.305 | 0.115 | 0.003 |
| Right | Control | Parietal | 1 | −0.298 | 0.114 | 0.003 |
| Left | Somatomotor | Somatomotor | 2 | −0.401 | 0.155 | 0.003 |
| Right | Control | Precuneus | 1 | −0.328 | 0.134 | 0.005 |
| Right | Control | Temporal | 1 | −0.306 | 0.125 | 0.005 |
| Right | Default | Posterior Cingulate Cortex | 1 | −0.338 | 0.139 | 0.005 |
| Right | Control | Parietal | 2 | −0.285 | 0.118 | 0.005 |
| Left | Default | Prefrontal Cortex | 10 | −0.264 | 0.111 | 0.006 |
| Right | Visual | Visual | 2 | −0.345 | 0.145 | 0.006 |
| Left | Default | Temporal | 8 | −0.293 | 0.124 | 0.006 |
| Left | Control | Parietal | 3 | −0.311 | 0.132 | 0.007 |
| Left | Default | Posterior Cingulate Cortex | 2 | −0.311 | 0.134 | 0.007 |
| Right | Dorsal Attention | Posterior | 3 | −0.298 | 0.128 | 0.007 |
| Left | Default | Parahippocampal Cortex | 1 | −0.293 | 0.127 | 0.008 |
| Left | Control | Lateral Prefrontal Cortex | 1 | −0.253 | 0.112 | 0.009 |
| Left | Default | Temporal | 5 | −0.338 | 0.150 | 0.009 |
| Left | Salience / Ventral Attention | Frontal Operculum | 2 | −0.246 | 0.109 | 0.009 |
| Right | Somatomotor | Somatomotor | 4 | −0.259 | 0.116 | 0.009 |
| Left | Default | Prefrontal Cortex | 1 | −0.243 | 0.110 | 0.011 |
| Right | Salience / Ventral Attention | Frontal Operculum | 3 | −0.255 | 0.115 | 0.011 |
| Left | Control | Lateral Prefrontal Cortex | 4 | −0.290 | 0.132 | 0.011 |
| Left | Control | Lateral Prefrontal Cortex | 5 | −0.299 | 0.137 | 0.011 |
| Right | Salience / Ventral Attention | Frontal Operculum | 1 | −0.219 | 0.100 | 0.011 |
| Left | Control | Parietal | 2 | −0.278 | 0.127 | 0.011 |
| Left | Default | Prefrontal Cortex | 13 | −0.271 | 0.125 | 0.012 |
| Right | Default | Posterior Cingulate Cortex | 2 | −0.283 | 0.131 | 0.012 |
| Left | Somatomotor | Somatomotor | 1 | −0.324 | 0.151 | 0.013 |
| Right | Dorsal Attention | Frontal Eye Fields | 2 | −0.265 | 0.125 | 0.014 |
| Left | Default | Posterior Cingulate Cortex | 1 | −0.276 | 0.132 | 0.015 |
| Left | Default | Posterior Cingulate Cortex | 3 | −0.216 | 0.103 | 0.015 |
| Right | Control | Ventral Prefrontal Cortex | 1 | −0.234 | 0.112 | 0.015 |
| Left | Control | Precuneus | 1 | −0.270 | 0.130 | 0.015 |
| Left | Default | Prefrontal Cortex | 9 | −0.257 | 0.125 | 0.015 |
| Right | Default | Parietal | 1 | −0.269 | 0.131 | 0.016 |
| Left | Default | Prefrontal Cortex | 11 | −0.228 | 0.113 | 0.018 |
| Left | Salience / Ventral Attention | Parietal Operculum | 1 | −0.275 | 0.136 | 0.018 |
| Left | Dorsal Attention | Posterior | 10 | −0.279 | 0.138 | 0.018 |
| Right | Visual | Visual | 11 | −0.272 | 0.135 | 0.018 |
| Right | Dorsal Attention | Posterior | 8 | −0.286 | 0.144 | 0.019 |
| Left | Default | Prefrontal Cortex | 3 | −0.248 | 0.126 | 0.021 |
| Right | Control | Medial Posterior Prefrontal Cortex | 1 | −0.233 | 0.119 | 0.021 |
| Right | Visual | Visual | 15 | −0.298 | 0.153 | 0.022 |
| Right | Somatomotor | Somatomotor | 1 | −0.300 | 0.156 | 0.024 |
| Left | Control | Lateral Prefrontal Cortex | 2 | −0.211 | 0.110 | 0.024 |
| Right | Dorsal Attention | Posterior | 10 | −0.245 | 0.129 | 0.026 |
| Right | Default | Medial Prefrontal Cortex | 7 | −0.211 | 0.113 | 0.028 |
| Left | Dorsal Attention | Posterior | 4 | −0.262 | 0.142 | 0.030 |
| Right | Default | Parietal | 3 | −0.228 | 0.124 | 0.032 |
| Right | Dorsal Attention | Posterior | 7 | −0.257 | 0.142 | 0.033 |
| Left | Default | Posterior Cingulate Cortex | 4 | −0.237 | 0.132 | 0.035 |
| Left | Default | Prefrontal Cortex | 5 | −0.249 | 0.139 | 0.035 |
| Right | Dorsal Attention | Posterior | 6 | −0.247 | 0.137 | 0.035 |
| Right | Default | Medial Prefrontal Cortex | 6 | −0.211 | 0.119 | 0.038 |
| Left | Dorsal Attention | Posterior | 9 | −0.241 | 0.138 | 0.041 |
| Right | Default | Ventral Prefrontal Cortex | 1 | −0.216 | 0.124 | 0.042 |
| Right | Control | Parietal | 3 | −0.220 | 0.127 | 0.043 |
| Left | Control | Temporal | 1 | −0.223 | 0.129 | 0.043 |
| Left | Default | Temporal | 3 | −0.263 | 0.152 | 0.043 |
| Left | Dorsal Attention | Frontal Eye Fields | 2 | −0.229 | 0.133 | 0.045 |
| Left | Salience / Ventral Attention | Medial | 3 | −0.198 | 0.115 | 0.045 |
| Left | Visual | Visual | 1 | −0.255 | 0.150 | 0.048 |

[1]Component names reflect the labels that were provided by Yeo et al. (2011). These labels split the seven networks into spatially connected components. We provide the component names and intra-component parcel numbers to aid in the identification of the parcels.

Note: The component names and the parcel numbers (within components) correspond to those that accompany the Schaefer et al., (2018) parcellation. Each row represents a single parcel in the brain. Component names reflect the component labels. More information about each parcel is available at their GitHub page (https://github.com/ThomasYeoLab/CBIG/tree/master/stable_projects/brain_parcellation/Schaefer2018_LocalGlobal). The quantity $\beta$ is the standardized regression coefficient and the quantity SE is the standard error.

Table S4. Results linking loneliness and neural responses: Cortical results (corresponding to Fig. 2a of the main manuscript)

Contrast: $ISC_{\{non\text{-}lonely,\ lonely\}} > ISC_{\{non\text{-}lonely,\ non\text{-}lonely\}}$

| Hemisphere | Network | Component Name[1] | Parcel Number | $\beta$ | SE | *p*-value (corrected) |
|---|---|---|---|---|---|---|
| Left | Control | Parietal | 1 | −0.393 | 0.111 | 0.000 |
| Left | Default | Parahippocampal Cortex | 1 | −0.465 | 0.128 | 0.000 |
| Right | Control | Parietal | 2 | −0.390 | 0.119 | 0.001 |
| Left | Control | Lateral Prefrontal Cortex | 3 | −0.314 | 0.100 | 0.001 |
| Left | Somatomotor | Somatomotor | 2 | −0.495 | 0.155 | 0.001 |
| Right | Dorsal Attention | Posterior | 4 | −0.414 | 0.129 | 0.001 |
| Right | Salience / Ventral Attention | Frontal Operculum | 2 | −0.377 | 0.121 | 0.001 |
| Right | Control | Parietal | 1 | −0.351 | 0.115 | 0.001 |
| Right | Visual | Visual | 2 | −0.442 | 0.146 | 0.001 |
| Left | Default | Temporal | 5 | −0.455 | 0.151 | 0.001 |
| Right | Control | Temporal | 1 | −0.372 | 0.126 | 0.001 |
| Left | Somatomotor | Somatomotor | 1 | −0.444 | 0.152 | 0.001 |
| Left | Control | Lateral Prefrontal Cortex | 6 | −0.368 | 0.127 | 0.001 |
| Right | Control | Lateral Prefrontal Cortex | 3 | −0.311 | 0.108 | 0.001 |
| Left | Control | Lateral Prefrontal Cortex | 1 | −0.323 | 0.113 | 0.001 |
| Right | Control | Precuneus | 1 | −0.385 | 0.135 | 0.001 |
| Left | Control | Parietal | 2 | −0.365 | 0.128 | 0.001 |
| Right | Dorsal Attention | Posterior | 8 | −0.407 | 0.144 | 0.001 |
| Right | Somatomotor | Somatomotor | 4 | −0.322 | 0.117 | 0.002 |
| Left | Limbic | Orbital Frontal Cortex | 1 | −0.254 | 0.094 | 0.002 |
| Left | Control | Lateral Prefrontal Cortex | 2 | −0.299 | 0.112 | 0.002 |
| Left | Dorsal Attention | Posterior | 7 | −0.383 | 0.144 | 0.003 |
| Left | Default | Prefrontal Cortex | 1 | −0.294 | 0.112 | 0.003 |
| Left | Control | Lateral Prefrontal Cortex | 5 | −0.360 | 0.138 | 0.003 |
| Left | Default | Temporal | 8 | −0.327 | 0.125 | 0.003 |
| Left | Salience / Ventral Attention | Frontal Operculum | 2 | −0.288 | 0.111 | 0.003 |
| Right | Visual | Visual | 11 | −0.352 | 0.135 | 0.003 |
| Right | Control | Lateral Prefrontal Cortex | 6 | −0.265 | 0.102 | 0.003 |
| Left | Default | Prefrontal Cortex | 8 | −0.253 | 0.099 | 0.003 |
| Right | Control | Medial Posterior Prefrontal Cortex | 3 | −0.248 | 0.099 | 0.004 |
| Left | Control | Parietal | 3 | −0.330 | 0.133 | 0.005 |
| Left | Control | Lateral Prefrontal Cortex | 4 | −0.328 | 0.133 | 0.005 |
| Left | Default | Prefrontal Cortex | 5 | −0.343 | 0.139 | 0.005 |
| Right | Somatomotor | Somatomotor | 2 | −0.386 | 0.156 | 0.005 |
| Right | Control | Medial Posterior Prefrontal Cortex | 1 | −0.292 | 0.120 | 0.005 |
| Right | Salience / Ventral Attention | Frontal Operculum | 3 | −0.280 | 0.117 | 0.006 |
| Right | Salience / Ventral Attention | Frontal Operculum | 1 | −0.241 | 0.102 | 0.006 |
| Right | Default | Ventral Prefrontal Cortex | 1 | −0.293 | 0.125 | 0.007 |
| Right | Somatomotor | Somatomotor | 1 | −0.363 | 0.157 | 0.007 |
| Left | Dorsal Attention | Posterior | 1 | −0.316 | 0.138 | 0.008 |
| Left | Default | Posterior Cingulate Cortex | 3 | −0.236 | 0.105 | 0.009 |
| Right | Default | Parietal | 1 | −0.296 | 0.132 | 0.009 |
| Left | Default | Prefrontal Cortex | 13 | −0.282 | 0.126 | 0.010 |
| Left | Default | Prefrontal Cortex | 12 | −0.238 | 0.108 | 0.011 |
| Right | Control | Parietal | 3 | −0.281 | 0.128 | 0.011 |
| Right | Visual | Visual | 15 | −0.336 | 0.153 | 0.011 |
| Left | Default | Prefrontal Cortex | 3 | −0.274 | 0.127 | 0.012 |
| Right | Visual | Visual | 1 | −0.296 | 0.140 | 0.013 |
| Left | Default | Temporal | 3 | −0.321 | 0.152 | 0.014 |
| Left | Control | Cingulate | 2 | −0.215 | 0.103 | 0.015 |
| Right | Control | Medial Posterior Prefrontal Cortex | 4 | −0.265 | 0.127 | 0.015 |
| Right | Default | Posterior Cingulate Cortex | 1 | −0.290 | 0.140 | 0.015 |
| Left | Control | Precuneus | 1 | −0.270 | 0.131 | 0.015 |
| Right | Default | Parietal | 3 | −0.258 | 0.125 | 0.015 |
| Left | Default | Posterior Cingulate Cortex | 4 | −0.273 | 0.133 | 0.016 |
| Left | Salience / Ventral Attention | Parietal Operculum | 3 | −0.234 | 0.116 | 0.018 |
| Right | Default | Medial Prefrontal Cortex | 7 | −0.228 | 0.114 | 0.019 |
| Left | Salience / Ventral Attention | Parietal Operculum | 1 | −0.272 | 0.137 | 0.019 |
| Right | Default | Posterior Cingulate Cortex | 2 | −0.263 | 0.132 | 0.019 |
| Left | Visual | Visual | 1 | −0.290 | 0.151 | 0.024 |
| Right | Control | Ventral Prefrontal Cortex | 1 | −0.215 | 0.114 | 0.027 |
| Right | Dorsal Attention | Posterior | 10 | −0.245 | 0.130 | 0.027 |
| Right | Dorsal Attention | Posterior | 3 | −0.241 | 0.129 | 0.029 |
| Left | Dorsal Attention | Frontal Eye Fields | 2 | −0.248 | 0.134 | 0.030 |
| Left | Default | Prefrontal Cortex | 10 | −0.206 | 0.112 | 0.032 |
| Left | Salience / Ventral Attention | Lateral Prefrontal Cortex | 1 | −0.197 | 0.109 | 0.035 |
| Left | Default | Prefrontal Cortex | 11 | −0.205 | 0.114 | 0.035 |
| Right | Control | Lateral Prefrontal Cortex | 7 | −0.221 | 0.123 | 0.035 |
| Left | Default | Posterior Cingulate Cortex | 1 | −0.239 | 0.133 | 0.035 |
| Left | Dorsal Attention | Posterior | 2 | −0.203 | 0.113 | 0.035 |
| Left | Default | Prefrontal Cortex | 9 | −0.219 | 0.126 | 0.042 |
| Right | Dorsal Attention | Posterior | 7 | −0.247 | 0.142 | 0.043 |
| Right | Dorsal Attention | Frontal Eye Fields | 2 | −0.215 | 0.126 | 0.047 |

[1]Component names reflect the labels that were provided by Yeo et al. (2011). These labels split the seven networks into spatially connected components. We provide the component names and intra-component parcel numbers to aid in the identification of the parcels.
Note: The component names and the parcel numbers (within components) correspond to those that accompany the Schaefer et al., (2018) parcellation. Each row represents a single parcel in the brain. Component names reflect the component labels. More information about each parcel is available at their GitHub page (https://github.com/ThomasYeoLab/CBIG/tree/master/stable_projects/brain_parcellation/Schaefer2018_LocalGlobal). The quantity $\beta$ is the standardized regression coefficient and the quantity SE is the standard error.

Table S5. Results linking loneliness and neural responses: Subcortical results (corresponding to Fig. 2b of the main manuscript)

Contrast: $ISC_{\{lonely, lonely\}} > ISC_{\{non-lonely, non-lonely\}}$

| Hemisphere | Brain Region | $\beta$ | SE | *p*-value (corrected) |
|---|---|---|---|---|
| Left | Nucleus Accumbens | −0.259 | 0.110 | 0.007 |
| Right | Pallidum | −0.154 | 0.083 | 0.029 |
| Left | Caudate Nucleus | −0.289 | 0.160 | 0.035 |

Table S6. Results linking loneliness and neural responses: Subcortical results (corresponding to Fig. 2b of the main manuscript)

Contrast: $ISC_{\{non-lonely, lonely\}} > ISC_{\{non-lonely, non-lonely\}}$

| Hemisphere | Brain Region | $\beta$ | SE | *p*-value (corrected) |
|---|---|---|---|---|
| Left | Caudate Nucleus | −0.203 | 0.090 | 0.009 |
| Left | Nucleus Accumbens | −0.153 | 0.071 | 0.012 |
| Right | Pallidum | −0.128 | 0.061 | 0.015 |
| Right | Nucleus Accumbens | −0.128 | 0.068 | 0.027 |

Table S7. Results linking loneliness and neural responses, controlling for objective social disconnection, demographic similarities, and friendships between participants: Cortical results (corresponding to Fig. 4a of the main manuscript)

Contrast: $ISC_{\{lonely, lonely\}} > ISC_{\{non-lonely, non-lonely\}}$

| Hemisphere | Network | Component Name[1] | Parcel Number | $\beta$ | SE | *p*-value (corrected) |
|---|---|---|---|---|---|---|
| Left | Control | Parietal | 1 | −0.684 | 0.215 | 0.002 |
| Right | Salience / Ventral Attention | Frontal Operculum | 2 | −0.701 | 0.241 | 0.003 |
| Left | Somatomotor | Somatomotor | 2 | −0.867 | 0.320 | 0.007 |
| Right | Control | Lateral Prefrontal Cortex | 3 | −0.576 | 0.218 | 0.007 |
| Right | Visual | Visual | 2 | −0.785 | 0.298 | 0.007 |
| Right | Dorsal Attention | Posterior | 4 | −0.663 | 0.257 | 0.008 |
| Right | Control | Lateral Prefrontal Cortex | 6 | −0.505 | 0.200 | 0.010 |
| Left | Control | Lateral Prefrontal Cortex | 1 | −0.568 | 0.226 | 0.010 |
| Right | Control | Parietal | 2 | −0.602 | 0.240 | 0.010 |
| Left | Control | Parietal | 2 | −0.619 | 0.254 | 0.011 |
| Left | Default | Temporal | 5 | −0.762 | 0.309 | 0.011 |
| Right | Control | Parietal | 1 | −0.577 | 0.236 | 0.011 |
| Right | Control | Temporal | 1 | −0.632 | 0.260 | 0.011 |
| Left | Default | Parahippocampal Cortex | 1 | −0.611 | 0.253 | 0.011 |
| Right | Somatomotor | Somatomotor | 4 | −0.541 | 0.230 | 0.014 |
| Left | Dorsal Attention | Posterior | 7 | −0.691 | 0.294 | 0.014 |
| Left | Control | Lateral Prefrontal Cortex | 5 | −0.640 | 0.276 | 0.014 |
| Left | Somatomotor | Somatomotor | 1 | −0.730 | 0.314 | 0.014 |
| Left | Limbic | Orbital Frontal Cortex | 1 | −0.396 | 0.173 | 0.016 |
| Left | Control | Parietal | 3 | −0.598 | 0.264 | 0.016 |
| Left | Control | Lateral Prefrontal Cortex | 3 | −0.436 | 0.192 | 0.016 |
| Left | Control | Lateral Prefrontal Cortex | 4 | −0.578 | 0.263 | 0.019 |
| Right | Salience / Ventral Attention | Frontal Operculum | 1 | −0.415 | 0.193 | 0.021 |
| Left | Default | Prefrontal Cortex | 1 | −0.442 | 0.210 | 0.025 |
| Left | Salience / Ventral Attention | Frontal Operculum | 2 | −0.435 | 0.208 | 0.025 |
| Right | Default | Parietal | 1 | −0.527 | 0.252 | 0.025 |
| Right | Dorsal Attention | Posterior | 3 | −0.547 | 0.261 | 0.025 |
| Left | Control | Lateral Prefrontal Cortex | 2 | −0.444 | 0.215 | 0.028 |
| Left | Default | Prefrontal Cortex | 12 | −0.432 | 0.211 | 0.029 |
| Right | Default | Ventral Prefrontal Cortex | 1 | −0.508 | 0.249 | 0.029 |
| Left | Default | Temporal | 8 | −0.502 | 0.247 | 0.029 |
| Right | Salience / Ventral Attention | Frontal Operculum | 3 | −0.462 | 0.228 | 0.030 |
| Left | Salience / Ventral Attention | Parietal Operculum | 3 | −0.476 | 0.237 | 0.031 |
| Left | Default | Posterior Cingulate Cortex | 3 | −0.405 | 0.204 | 0.033 |
| Right | Visual | Visual | 11 | −0.526 | 0.266 | 0.033 |
| Left | Default | Prefrontal Cortex | 5 | −0.564 | 0.286 | 0.033 |
| Right | Somatomotor | Somatomotor | 1 | −0.632 | 0.320 | 0.033 |
| Left | Control | Lateral Prefrontal Cortex | 6 | −0.497 | 0.260 | 0.041 |

[1]Component names reflect the labels that were provided by Yeo et al. (2011). These labels split the seven networks into spatially connected components. We provide the component names and intra-component parcel numbers to aid in the identification of the parcels.
Note: The component names and the parcel numbers (within components) correspond to those that accompany the Schaefer et al., (2018) parcellation. Each row represents a single parcel in the brain. Component names reflect the component labels. More information about each parcel is available at their GitHub page (https://github.com/ThomasYeoLab/CBIG/tree/master/stable_projects/brain_parcellation/Schaefer2018_LocalGlobal). The quantity *β* is the standardized regression coefficient and the quantity SE is the standard error.

Table S8. Results linking loneliness and neural responses, controlling for objective social disconnection, demographic similarities, and friendships between participants: Subcortical results (corresponding to Fig. 4b of the main manuscript)

Contrast: $ISC_{\{lonely, lonely\}} > ISC_{\{non-lonely, non-lonely\}}$

| Hemisphere | Brain Region | *β* | SE | *p*-value (corrected) |
|---|---|---|---|---|
| Left | Nucleus Accumbens | −0.205 | 0.106 | 0.037 |

Table S9. Results linking loneliness and neural responses, controlling for objective social disconnection, demographic similarities, and friendships between participants: Cortical results (corresponding to Fig. 4a of the main manuscript)

Contrast: $ISC_{\{lonely, lonely\}} > ISC_{\{non-lonely, lonely\}}$

| Hemisphere | Network | Component Name[1] | Parcel Number | *β* | SE | *p*-value (corrected) |
|---|---|---|---|---|---|---|
| Left | Control | Parietal | 1 | −0.321 | 0.113 | 0.004 |
| Right | Salience / Ventral Attention | Frontal Operculum | 2 | −0.323 | 0.125 | 0.008 |
| Left | Dorsal Attention | Posterior | 7 | −0.357 | 0.149 | 0.012 |
| Right | Control | Lateral Prefrontal Cortex | 3 | −0.275 | 0.115 | 0.012 |
| Right | Dorsal Attention | Posterior | 4 | −0.317 | 0.132 | 0.012 |
| Left | Somatomotor | Somatomotor | 2 | −0.384 | 0.161 | 0.012 |
| Right | Control | Lateral Prefrontal Cortex | 6 | −0.244 | 0.107 | 0.016 |
| Right | Dorsal Attention | Posterior | 3 | −0.304 | 0.134 | 0.016 |
| Right | Visual | Visual | 2 | −0.339 | 0.151 | 0.017 |
| Left | Control | Parietal | 3 | −0.301 | 0.135 | 0.017 |
| Left | Salience / Ventral Attention | Parietal Operculum | 3 | −0.271 | 0.123 | 0.019 |
| Right | Control | Parietal | 1 | −0.270 | 0.123 | 0.019 |
| Left | Default | Prefrontal Cortex | 12 | −0.241 | 0.112 | 0.021 |
| Right | Control | Temporal | 1 | −0.288 | 0.133 | 0.021 |
| Left | Control | Lateral Prefrontal Cortex | 1 | −0.250 | 0.118 | 0.025 |
| Left | Control | Parietal | 2 | −0.274 | 0.131 | 0.025 |
| Left | Control | Lateral Prefrontal Cortex | 5 | −0.294 | 0.141 | 0.025 |
| Right | Somatomotor | Somatomotor | 4 | −0.246 | 0.120 | 0.029 |
| Right | Dorsal Attention | Frontal Eye Fields | 2 | −0.266 | 0.131 | 0.029 |
| Left | Control | Lateral Prefrontal Cortex | 4 | −0.274 | 0.135 | 0.029 |
| Right | Control | Parietal | 2 | −0.250 | 0.125 | 0.031 |
| Left | Default | Prefrontal Cortex | 10 | −0.228 | 0.115 | 0.033 |
| Right | Default | Parietal | 1 | −0.255 | 0.130 | 0.035 |
| Left | Default | Temporal | 5 | −0.305 | 0.156 | 0.035 |
| Left | Default | Temporal | 8 | −0.249 | 0.128 | 0.035 |
| Left | Somatomotor | Somatomotor | 1 | −0.310 | 0.159 | 0.035 |
| Right | Salience / Ventral Attention | Frontal Operculum | 3 | −0.231 | 0.119 | 0.037 |

[1]Component names reflect the labels that were provided by Yeo et al. (2011). These labels split the seven networks into spatially connected components. We provide the component names and intra-component parcel numbers to aid in the identification of the parcels.

Note: The component names and the parcel numbers (within components) correspond to those that accompany the Schaefer et al., (2018) parcellation. Each row represents a single parcel in the brain. Component names reflect the

component labels. More information about each parcel is available at their GitHub page (https://github.com/ThomasYeoLab/CBIG/tree/master/stable_projects/brain_parcellation/Schaefer2018_LocalGlobal). The quantity $\beta$ is the standardized regression coefficient and the quantity SE is the standard error.

Table S10. Results linking loneliness and neural responses, controlling for objective social disconnection, demographic similarities, and friendships between participants: Cortical results (corresponding to Fig. 4a of the main manuscript)

Contrast: $ISC_{\{non\text{-}lonely,\ lonely\}} > ISC_{\{non\text{-}lonely,\ non\text{-}lonely\}}$

| Hemisphere | Network | Component Name[1] | Parcel Number | $\beta$ | SE | *p*-value (corrected) |
|---|---|---|---|---|---|---|
| Left | Control | Parietal | 1 | −0.363 | 0.113 | 0.002 |
| Left | Default | Parahippocampal Cortex | 1 | −0.396 | 0.130 | 0.003 |
| Left | Somatomotor | Somatomotor | 2 | −0.484 | 0.161 | 0.003 |
| Right | Salience / Ventral Attention | Frontal Operculum | 2 | −0.378 | 0.125 | 0.003 |
| Left | Default | Temporal | 5 | −0.456 | 0.156 | 0.003 |
| Right | Visual | Visual | 2 | −0.447 | 0.151 | 0.003 |
| Left | Control | Lateral Prefrontal Cortex | 3 | −0.296 | 0.104 | 0.004 |
| Right | Control | Parietal | 2 | −0.352 | 0.125 | 0.004 |
| Left | Control | Parietal | 2 | −0.345 | 0.131 | 0.007 |
| Left | Control | Lateral Prefrontal Cortex | 1 | −0.318 | 0.119 | 0.007 |
| Left | Limbic | Orbital Frontal Cortex | 1 | −0.254 | 0.096 | 0.007 |
| Left | Somatomotor | Somatomotor | 1 | −0.420 | 0.159 | 0.007 |
| Right | Control | Lateral Prefrontal Cortex | 3 | −0.301 | 0.115 | 0.007 |
| Right | Dorsal Attention | Posterior | 4 | −0.345 | 0.132 | 0.007 |
| Right | Control | Temporal | 1 | −0.344 | 0.134 | 0.008 |
| Right | Control | Parietal | 1 | −0.307 | 0.123 | 0.010 |
| Left | Control | Lateral Prefrontal Cortex | 5 | −0.346 | 0.141 | 0.011 |
| Left | Control | Lateral Prefrontal Cortex | 6 | −0.326 | 0.134 | 0.011 |
| Right | Control | Lateral Prefrontal Cortex | 6 | −0.260 | 0.107 | 0.011 |
| Right | Somatomotor | Somatomotor | 4 | −0.295 | 0.120 | 0.011 |
| Left | Default | Prefrontal Cortex | 1 | −0.261 | 0.112 | 0.014 |
| Left | Default | Prefrontal Cortex | 5 | −0.339 | 0.146 | 0.014 |
| Left | Default | Prefrontal Cortex | 8 | −0.229 | 0.099 | 0.014 |
| Right | Somatomotor | Somatomotor | 2 | −0.375 | 0.162 | 0.014 |
| Left | Control | Lateral Prefrontal Cortex | 4 | −0.304 | 0.135 | 0.017 |
| Left | Dorsal Attention | Posterior | 7 | −0.334 | 0.149 | 0.017 |
| Left | Salience / Ventral Attention | Frontal Operculum | 2 | −0.248 | 0.111 | 0.017 |
| Right | Default | Ventral Prefrontal Cortex | 1 | −0.288 | 0.129 | 0.017 |
| Right | Salience / Ventral Attention | Frontal Operculum | 1 | −0.233 | 0.104 | 0.017 |
| Right | Visual | Visual | 11 | −0.302 | 0.136 | 0.018 |
| Left | Control | Parietal | 3 | −0.297 | 0.136 | 0.019 |
| Right | Somatomotor | Somatomotor | 1 | −0.353 | 0.161 | 0.019 |
| Left | Control | Lateral Prefrontal Cortex | 2 | −0.248 | 0.114 | 0.019 |
| Right | Control | Medial Posterior Prefrontal Cortex | 3 | −0.213 | 0.101 | 0.025 |
| Right | Default | Parietal | 1 | −0.272 | 0.130 | 0.025 |
| Right | Control | Parietal | 3 | −0.271 | 0.131 | 0.027 |
| Right | Dorsal Attention | Posterior | 8 | −0.305 | 0.149 | 0.029 |
| Left | Default | Posterior Cingulate Cortex | 3 | −0.219 | 0.109 | 0.031 |
| Right | Control | Precuneus | 1 | −0.272 | 0.137 | 0.033 |
| Left | Default | Temporal | 8 | −0.253 | 0.128 | 0.033 |
| Right | Salience / Ventral Attention | Frontal Operculum | 3 | −0.231 | 0.120 | 0.038 |
| Right | Control | Medial Posterior Prefrontal Cortex | 1 | −0.230 | 0.121 | 0.042 |
| Right | Default | Medial Prefrontal Cortex | 7 | −0.222 | 0.119 | 0.047 |

[1]Component names reflect the labels that were provided by Yeo et al. (2011). These labels split the seven networks into spatially connected components. We provide the component names and intra-component parcel numbers to aid in the identification of the parcels.

Note: The component names and the parcel numbers (within components) correspond to those that accompany the Schaefer et al., (2018) parcellation. Each row represents a single parcel in the brain. Component names reflect the component labels. More information about each parcel is available at their GitHub page (https://github.com/ThomasYeoLab/CBIG/tree/master/stable_projects/brain_parcellation/Schaefer2018_LocalGlobal). The quantity $\beta$ is the standardized regression coefficient and the quantity SE is the standard error.

Table S11. Results linking loneliness and neural responses, controlling for objective social disconnection, demographic similarities, and friendships between participants: Subcortical results (corresponding to Fig. 4b of the main manuscript)

Contrast: $ISC_{\{non\text{-}lonely,\ lonely\}} > ISC_{\{non\text{-}lonely,\ non\text{-}lonely\}}$

| Hemisphere | Brain Region | $\beta$ | SE | *p*-value (corrected) |
|---|---|---|---|---|
| Right | Pallidum | −0.142 | 0.062 | 0.016 |
| Left | Caudate Nucleus | −0.195 | 0.093 | 0.025 |

## Supplementary results: Objective social disconnection

**Objective social disconnection**. As we noted in the main manuscript, we used out-degree centrality as a measure of objective social disconnection. The out-degree centrality of the participants ranged from 0 to 23 (with mean = 4.73, median = 3, and SD = 5.153).

We split the participants into two groups based on a median split of their objective social-disconnection measure. We categorized participants into the low objective social-disconnection group if they had an out-degree that was larger than the median (specifically, if it was more than 3; there were $n_{\text{low}} = 25$ such people) and into the high objective social-disconnection group if they had an out-degree that was less than or equal to the median (specifically, if it was less than or equal to 3; there were $n_{\text{high}} = 38$ such people). We then transformed the participant-level variable into a dyad-level variable. We categorized the dyads into (1) {high, high} if both individuals in a dyad had a high out-degree centrality, (2) {low, low} if both individuals in a dyad had a low out-degree centrality, and (3) {low, high} if one individual in a dyad had a low out-degree centrality and the other individual had a high out-degree centrality. Of the 1,952 unique dyads with complete fMRI and social-network data, 300 dyads were {high, high} dyads, 702 were {low, high} dyads, and 950 were {low, low} dyads with respect to their levels of objective social disconnection.

**Relating subjective and objective social disconnection**. We found that subjective and objective social disconnection were significantly correlated with one another, such that greater loneliness was associated significantly with a smaller number of friends (with a Pearson correlation coefficient of $r(61) = -0.337$ and a $p$-value of $p = 0.007$).

## Supplementary results: Controlling for in-degree centrality

**Results associating neural similarity with binarized loneliness when we control for in-degree centrality and self-reported demographic traits.** For exploratory purposes, we fit analogous models to those that we described in the main manuscript in relating loneliness with neural similarity when we control for objective social disconnection, friendships between participants, and demographic similarities of the Results section, except that we controlled for individuals' in-degree centralities instead of their out-degree centralities. We calculated the in-degree centrality of each individual as follows. As we noted in the main manuscript (see Characterizing subjective and objective social disconnection in the Method section), we characterized the social networks of individuals who lived in two different residential communities of first-year students at a large state university in the United States. Using the responses of the individuals, we constructed a directed network for each of the two communities. In each of these networks, a node represents an individual and a directed edge represents one individual nominating another as a friend. For each individual, we calculated in-degree centrality, which counts the number of times that the individual was nominated as a friend by others in the network. We then used a median split of the in-degree centralities to binarize our sample into high-centrality and low-centrality groups. This choice is consistent with recent studies that related neural similarity with behavioral measures (Finn et al., 2018; Leong et al., 2020). We classified participants as part of the high-centrality group if they had an in-degree centrality that was larger than the median (specifically, if it was more than 2; there were $n_{\text{high}}$ = 23 such people) and into the low-centrality group if they had an in-degree centrality that was less than or equal to the median (specifically, if it was less than or equal to 2; there were $n_{\text{low}}$ = 40 such people). We then transformed the individual-level binarized in-degree centrality measure

into a dyad-level variable. We categorized the dyads into (1) {high, high} if both individuals in a dyad had a high in-degree centrality, (2) {low, low} if both individuals in a dyad had a low in-degree centrality, and (3) {low, high} if one individual in a dyad had a low in-degree centrality and the other individual had a high in-degree centrality.

We fit analogous models as the ones that we described in “Relating loneliness with neural similarity when we control for objective social disconnection, friendships between participants, and demographic similarities” in the “Results” section of the main manuscript. Specifically, for each brain region, we fit linear mixed-effects models, with the ISCs in the brain region as the dependent variable and the dyad-level loneliness variable as the independent variable of interest, while controlling for in-degree centrality, friendships between individuals in the dyad, and dyadic similarities in age, gender, ethnicity, and home country as covariates of no interest. The results of these calculations (see Fig. S3) are similar to those that we reported in the main manuscript (see Figs. 2 and 4).

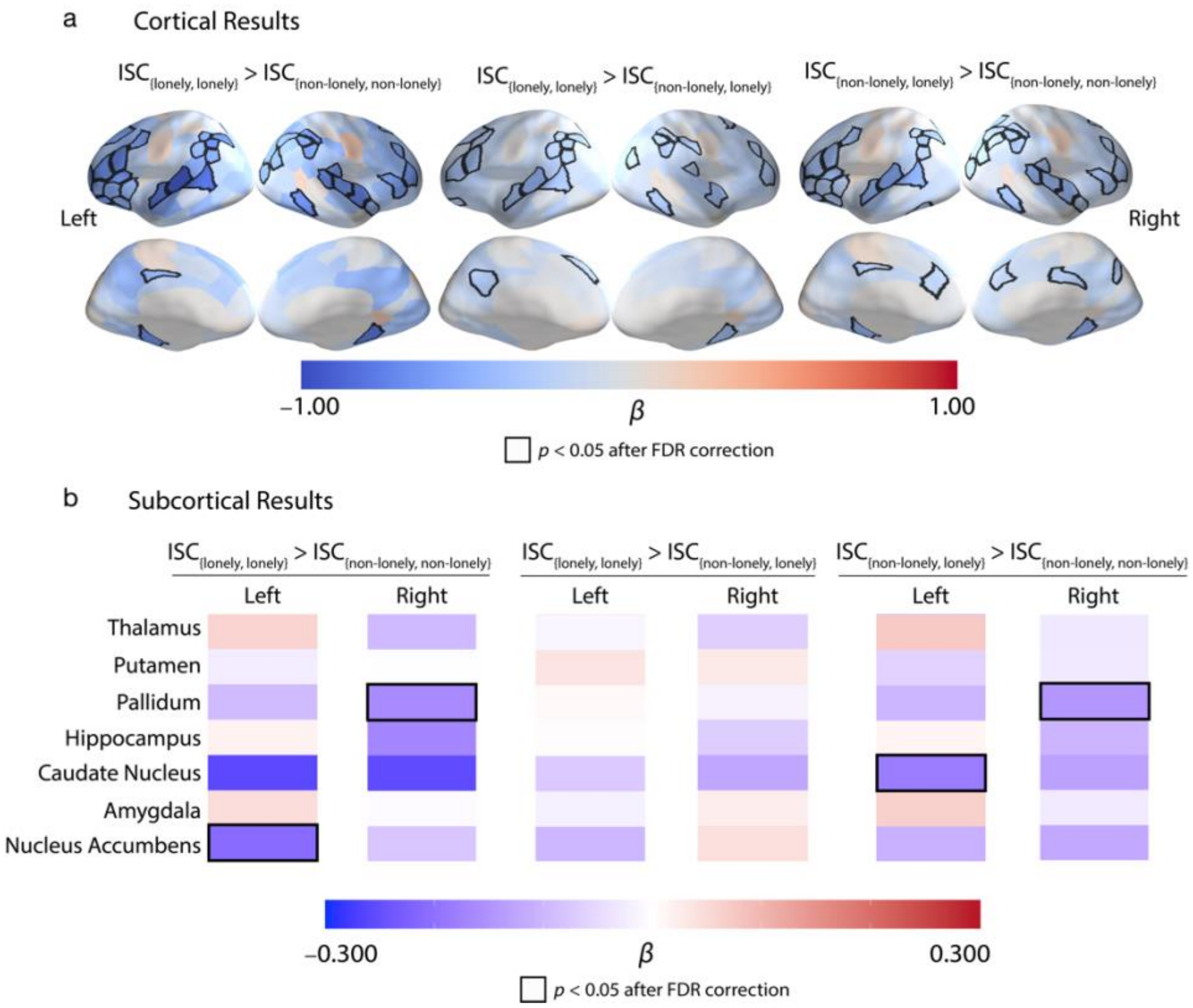

**Fig. S3.** Linking loneliness to idiosyncratic neural responses while controlling for in-degree centrality, demographic similarities, and friendships between participants. **(a)** As in our results in the main manuscript, we observed smaller ISCs in brain regions (including the VLPFC, DLPFC, STS, IPL, and SPL) that are associated with social cognition, shared understanding of events, and friendship in dyads with individuals who were both lonely (i.e., {lonely, lonely}) than in dyads with individuals who were both non-lonely (i.e., {non-lonely, non-lonely}). We observed similar patterns when we compared dyads with two lonely individuals (i.e., {lonely, lonely}) with dyads with one non-lonely individual and one lonely individual (i.e., {non-lonely, lonely}) and when we compared dyads with one non-lonely individual and one lonely individual (i.e., {non-lonely, lonely}) to dyads with two non-lonely individuals (i.e., {non-lonely, non-lonely}). **(b)** The ISCs were smaller in the left nucleus accumbens and right pallidum in dyads with two lonely individuals than in dyads with two non-lonely individuals. The labels “Left” and “Right” refer to the hemispheres of the brain regions that are listed in the left panel. The quantity $\beta$ is the standardized regression coefficient. Regions with significant associations between loneliness and ISC are outlined in black (using an FDR-corrected significance threshold of $p < 0.05$).

## Supplementary discussion

We measured objective social disconnection (i.e., number of friends) 2–3 months after we collected our neuroimaging and loneliness data, and all data were collected during a time period (the first half of the participants' first year of college) during which both objective and subjective social disconnection may (1) be impacted strongly by situational factors (e.g., whether or not one just moved far away from home) and (2) be subject to greater fluctuations than under other circumstances. Future studies may benefit from simultaneously collecting objective social disconnection, loneliness, and neural data on more established communities. Employing such an approach may also confer greater sensitivity to detect relationships between the relative levels of loneliness of study participants and how idiosyncratically they process the world around them.